\documentclass{jpp}

\usepackage{graphicx}

\usepackage[utf8]{inputenc}
\usepackage[T1]{fontenc}
\usepackage{amsmath}

\usepackage{bm,xcolor}
\usepackage{tikz}
\usetikzlibrary{spy}
\usepackage{standalone}
\usepackage{appendix}
\usepackage{placeins}
\usepackage{booktabs}
\usepackage{siunitx}
\usepackage{url}
\usepackage{array}
\usepackage{amsmath}           
\usepackage{amssymb}
\usepackage{pgfplots}
    \usepgfplotslibrary{colorbrewer}
\usepackage{pgfplotstable}
\pgfplotsset{compat=1.16}
\usepgfplotslibrary{groupplots}

\shorttitle{Optimization for quasi-symmetry on surfaces}
\shortauthor{A. Giuliani, F. Wechsung, M. Landreman, G. Stadler, A. Cerfon}

\title{Direct computation of magnetic surfaces in Boozer coordinates and coil optimization for quasi-symmetry}

\author{Andrew Giuliani\aff{1}
  \corresp{\email{giuliani@cims.nyu.edu}},
  Florian Wechsung\aff{1}, Georg Stadler\aff{1}, Antoine Cerfon\aff{1}, \and Matt Landreman\aff{2} }

\affiliation{\aff{1}Courant Institute of Mathematical Sciences, New York University,
New York, NY, USA \aff{2} University of Maryland-College Park, MD, USA}

\newcommand{\nfp}{n_{\text{fp}}}
\newcommand{\ntor}{n_{\text{tor}}}
\newcommand{\mpol}{m_{\text{pol}}}
\begin{document}

\maketitle

\begin{abstract}
We propose a new method to compute magnetic surfaces that are parametrized in
Boozer coordinates for vacuum magnetic fields. We also propose a measure for
quasi-symmetry on the computed surfaces and use it to design coils that generate
a magnetic field that is quasi-symmetric on those surfaces. 
The rotational transform of the field and
complexity measures for the coils are also controlled in the design problem.
Using an adjoint approach, we are able to obtain analytic derivatives for this
optimization problem, yielding an efficient gradient-based algorithm.
Starting from an initial coil set that presents nested magnetic surfaces for a large fraction of the
volume, our method converges rapidly to coil systems generating fields with excellent quasi-symmetry and low particle losses.
In particular for low complexity coils, we are able to
significantly improve the performance compared to coils obtained from the standard
two-stage approach, e.g.~reduce  losses of fusion-produced alpha particles born at half-radius from $17.7\%$ to $6.6\%$.
We also demonstrate 16-coil configurations with alpha loss < $1\%$ and neoclassical transport magnitude $\epsilon_{\mathrm{eff}}^{3/2}$ less than approximately $5\times 10^{-9}.$
\end{abstract}

\section{Introduction}
Single-stage coil design, in which one optimizes for the geometry and currents of electromagnetic coils at the same time as the target magnetic configuration, is a promising approach for stellarator design as it considers physics goals and engineering constraints simultaneously \citep{Giuliani2020Singlestage,henneberg2021combined,Wechsung2021Singlestage,yu2022existence}. It has the potential to yield designs with coil systems that are easier to manufacture and position. In the single-stage method, the optimization objective must explicitly favor both the existence of nested flux surfaces for a large fraction of the plasma volume, as well as good particle confinement.
This is empirically known to be more complicated than in the more common two-stage coil design strategy \citep{Merkel_1987,drevlak_1998,strickler_2002,strickler_2004,brown_2015,Zhu2017,Landreman_2017,Paul_2018,Singh_2020}.
One single-stage optimization approach was demonstrated in~\cite{Giuliani2020Singlestage},
based on the near-axis expansion for quasi-symmetry in \cite{garren1991,landreman_2018,landreman_2019,LandremanSengupta2019}.
This approach was motivated by the fact that quasi-symmetric magnetic fields are known to have good confinement properties~\citep{helander2014theory,LandremanPrecise}. 
We demonstrated that this single-stage method effectively produced vacuum quasi-symmetric magnetic fields in the vicinity of the magnetic axis. However, we also observed that quasi-symmetry degraded with distance from the magnetic axis. The purpose of the present article is to introduce an optimization objective that targets quasi-symmetry at an arbitrary number of surfaces, potentially far from the magnetic axis. We show that this formulation results in coil systems producing a vacuum field with excellent quasi-symmetry for a large fraction of the plasma volume.

In order to optimize for quasi-symmetry on a given surface, we need to measure the degree of departure of the magnetic field from quasi-symmetry, which we do using Boozer coordinates \citep{BoozerCoordinates}. The first contribution of the present article thus is a new numerical method for computing magnetic surfaces directly parametrized in Boozer coordinates. It relies on a partial differential equation (PDE) directly derived from to contra- and covariant expressions for the field in Boozer coordinates. This PDE can be discretized either as a balanced or over-determined system of nonlinear equations, solved exactly or in a least squares sense, respectively. 

The second contribution of this article is the formulation of a single-stage optimization problem for highly accurate quasi-symmetry and a target rotational transform on an arbitrary number of surfaces subject to engineering constraints for the coils. Penalty terms 
are constructed to favor simpler coils by constraining their minimum pairwise distance, length, maximum curvature, and mean squared curvature, similar to constraints used
for two-stage coil design \citep{Zhu2017,kruger_zhu_bader_et_al_21,Singh_2020}.  
We consider zero-thickness filament coils for simplicity, but our approach can be generalized to coils of finite thickness, e.g., following the method proposed in \cite{Singh_2020}.

We rely on gradient-based optimization tools for efficient and robust convergence to minimizers. To compute the required gradients in an efficient and accurate manner, we  use an adjoint sensitivity approach. As a result, the performance of the method to compute gradients is independent of the number of coil degrees of freedom, which is of the order of hundreds in our example.

We apply our method to improve a set of coils obtained from the classic two-stage approach. These coils were optimized to approximate the precise quasi-symmetric magnetic fields recently discovered in \cite{LandremanPrecise}. Our approach allows us to improve quasi-symmetry on surfaces compared to the results from the two-stage approach, without increasing our measures for coil complexity, i.e., maximum curvature, mean squared curvature, minimum pairwise coil distance, and total coil length.
For this particular example, our approach can be viewed either as a third stage improving an existing design found from the standard two-stage approach, or as single stage coil optimization approach using a good initialization for the coils.

The resulting configurations have impressive confinement, even though they only rely on 16 coils. When the three longest coil configurations are scaled to a reactor, losses of collisionless fusion-produced alpha particles born at half-radius are found to be below $1\%$. Moreover, the magnitude of the 3D neoclassical transport metric $\epsilon_{\mathrm{eff}}^{3/2}$ \citep{NEO} is less than approximately $5 \times 10^{-9}$. To our knowledge this value is smaller than in any previous stellarator, smaller even than in tokamaks (Figure 15 of \cite{Spong2015}). 

The structure of this article is as follows. Sections 2 and 3 focus on our new numerical method for computing surfaces in Boozer coordinates: in Section 2, we introduce a parametrization to approximate toroidal surfaces, and in Section 3 we introduce the partial differential equations that the surfaces in Boozer coordinates satisfy, and the numerical method used to solve them.  In Section 4, we define the coil optimization problem, which combines our surface computation approach, the physics goals of quasi-symmetry and a target rotational transform, with standard engineering constraints. In Section 5, we study the performance of our new optimization scheme for the design of a quasi-axisymmetric magnetic configuration, and we summarize our work in Section 6.

\section{Surface parametrization}\label{sec:surface-param}
We approximate a generic toroidal surface $\bm \Sigma$ that satisfies $n_{\mathrm{fp}}$-rotational symmetry with a finite-dimensional surface
$\bm\Sigma_s(\varphi,\theta): [0,1)^2 \rightarrow (x,y,z)$. The parametrization of  $\bm\Sigma_s(\varphi,\theta)$ detailed below directly incorporates the discrete rotational symmetry, making it efficient and convenient to work with. For $m_{\text{pol}}, n_{\text{tor}}\ge 0$, it is given by  
\begin{align*}
    x &= \cos(2\pi\varphi) \hat x - \sin(2\pi\varphi) \hat y \\
    y &= \sin(2\pi\varphi) \hat x + \cos(2\pi\varphi) \hat y \\
    z &= \sum^{ 2 m_{\text{pol}} }_{i = 0} \sum^{2 n_{\text{tor}} }_{j = 0} z_{i,j} w_i(\theta)v_j(\varphi)
\end{align*}
with
\begin{align*}
\hat x =  \sum^{ 2 m_{\text{pol}} }_{i = 0} \sum^{2 n_{\text{tor}} }_{j = 0} x_{i,j} w_i(\theta)v_j(\varphi) \quad \text{ and } \quad \hat y &=  \sum^{ 2 m_{\text{pol}} }_{i = 0} \sum^{2 n_{\text{tor}} }_{j = 0} y_{i,j} w_i(\theta)v_j(\varphi)
\end{align*}
and the basis functions
\begin{align*}
    w(\theta) &= (1, \cos(2 \pi \theta), \sin(2 \pi \theta),  \hdots, \cos(2 \pi  m_{\text{pol}}\theta), \sin(2 \pi m_{\text{pol}} \theta) ) \\
    v(\varphi) &= (1, \cos(2 \pi n_{\text{fp}} \varphi), \sin(2\pi n_{\text{fp}}\varphi),  \hdots, \cos(2 \pi n_{\text{fp}} n_{\text{tor}}\theta), \sin(2 \pi n_{\text{fp}} n_{\text{tor}} \theta)). 
\end{align*}
The functions $\hat x$, $\hat y$ and $z$ are linear combinations of tensor-product basis functions, which are $n_{\text{fp}}$-periodic in $\varphi$.
 In order to achieve rotational symmetry around the $\{x=y=0\}$ axis, we apply a rotation to $\hat x$ and $\hat y$, which ensures that $\bm\Sigma_s$ is $n_{\text{fp}}$-rotational symmetric, as we show explicitly in Appendix \ref{SurfaceRot}.

For stellarator-symmetric surfaces, we additionally require that  
\begin{equation} \label{eq:ss}
\begin{aligned}
    x(\varphi,\theta) &= x(-\varphi,-\theta) \\
    y(\varphi,\theta) &= -y(-\varphi,-\theta)\\
    z(\varphi,\theta) &= -z(-\varphi,-\theta).
\end{aligned}
\end{equation}
This is achieved by removing the basis functions in the $x$, $y$, $z$ expansions that do not satisfy 
\eqref{eq:ss}.
In particular, basis functions of the form $\sin(2m\pi n_{\mathrm{fp}}\varphi)\cos(2n\pi \theta)$ and $\cos(2m\pi n_{\mathrm{fp}}\varphi)\sin(2n\pi \theta)$ are removed from $\hat x$.
Similarly, basis functions of the form 1, $\sin(2m\pi n_{\mathrm{fp}}\varphi)\sin(2n\pi \theta)$, and $\cos(2m n_{\mathrm{fp}}\pi \varphi)\cos(2n\pi \theta)$ are removed from $\hat y$ and $z$.
This results in a surface representation that has $n_s=3[2n_{\text{tor}}m_{\text{pol}} + m_{\text{pol}} + n_{\text{tor}} + 1] - 2$ degrees of freedom. In the remainder of this manuscript we only consider stellarator-symmetric surfaces.
Note that the work described below is not fundamentally restricted to this surface parametrization and others can be used.
For example, we have also used a surface parametrization where the geometric degrees of freedom correspond to pointwise solution values and spatial derivatives are obtained using spectral differentiation matrices.

Note that the surface parametrization used, e.g., in VMEC \citep{vmec}
 uses a parameterization in cylindrical coordinates $\bm\Sigma_s(\phi,
\theta) = (r(\phi, \theta),\phi, z(\phi,\theta))$. This
parameterization has fewer degrees of freedom to represent surfaces,
but it is inappropriate for our purposes because it cannot directly be
used to parametrize a surface in Boozer coordinates. This is because
the toroidal angle used is the cylindrical angle $\phi$, which does
not correspond to the Boozer toroidal angle $\varphi$.  It is not the choice of 
coordinate system that is the issue as either cylindrical or Cartesian coordinates
could describe a surface; we cannot directly use the VMEC representation 
due to the incompatibility of Boozer $\varphi$ and cylindrical
$\phi$ as described above.

\section{Computing surfaces}\label{sec:surfaces}
\begin{figure}
\centering
    \begin{tikzpicture}
    \node (A) at (0,0) {   \begin{tikzpicture}
    \def\thickness{2};
    \draw[help lines, color=gray!30, dashed] (0,0) grid (7.1,7.1);
    \draw[->,ultra thick] (0,0)--(7.5,0) node[right]{\Large $\varphi$};
    \draw[->,ultra thick] (0,0)--(0,7.5) node[above]{\Large $\theta$};
    \draw[ultra thick] (7,-0.25)--(7,0.25);
    \node at (7,-0.75) {\Large $1/n_{\text{fp}}$};
    \draw[ultra thick] (-0.25,7)--(0.25,7);
    \node at (-0.5,7) {\Large $1$};

    \node at (0,0) [circle,fill,inner sep=\thickness pt]{};
    \node at (0,1) [circle,fill,inner sep=\thickness pt]{};
    \node at (0,2) [circle,fill,inner sep=\thickness pt]{};
    \node at (0,3) [circle,fill,inner sep=\thickness pt]{};
    \node at (0,4) [circle,fill,inner sep=\thickness pt]{};
    \node at (0,5) [circle,fill,inner sep=\thickness pt]{};
    \node at (0,6) [circle,fill,inner sep=\thickness pt]{};
  
    \node at (1,0) [circle,fill,inner sep=\thickness pt]{};
    \node at (2,0) [circle,fill,inner sep=\thickness pt]{};
    \node at (3,0) [circle,fill,inner sep=\thickness pt]{};
    \node at (4,0) [circle,fill,inner sep=\thickness pt]{};
    \node at (5,0) [circle,fill,inner sep=\thickness pt]{};
    \node at (6,0) [circle,fill,inner sep=\thickness pt]{};
    
    \node at (1,1) [circle,fill,inner sep=\thickness pt]{};
    \node at (2,1) [circle,fill,inner sep=\thickness pt]{};
    \node at (3,1) [circle,fill,inner sep=\thickness pt]{};
    \node at (4,1) [circle,fill,inner sep=\thickness pt]{};
    \node at (5,1) [circle,fill,inner sep=\thickness pt]{};
    \node at (6,1) [circle,fill,inner sep=\thickness pt]{};
    
    \node at (1,2) [circle,fill,inner sep=\thickness pt]{};
    \node at (2,2) [circle,fill,inner sep=\thickness pt]{};
    \node at (3,2) [circle,fill,inner sep=\thickness pt]{};
    \node at (4,2) [circle,fill,inner sep=\thickness pt]{};
    \node at (5,2) [circle,fill,inner sep=\thickness pt]{};
    \node at (6,2) [circle,fill,inner sep=\thickness pt]{};
    
    \node at (1,3) [circle,fill,inner sep=\thickness pt]{};
    \node at (2,3) [circle,fill,inner sep=\thickness pt]{};
    \node at (3,3) [circle,fill,inner sep=\thickness pt]{};
    \node at (4,3) [circle,fill,inner sep=\thickness pt]{};
    \node at (5,3) [circle,fill,inner sep=\thickness pt]{};
    \node at (6,3) [circle,fill,inner sep=\thickness pt]{};
    
    \node at (1,4) [circle,fill,inner sep=\thickness pt]{};
    \node at (2,4) [circle,fill,inner sep=\thickness pt]{};
    \node at (3,4) [circle,fill,inner sep=\thickness pt]{};
    \node at (4,4) [circle,fill,inner sep=\thickness pt]{};
    \node at (5,4) [circle,fill,inner sep=\thickness pt]{};
    \node at (6,4) [circle,fill,inner sep=\thickness pt]{};
    
    \node at (1,5) [circle,fill,inner sep=\thickness pt]{};
    \node at (2,5) [circle,fill,inner sep=\thickness pt]{};
    \node at (3,5) [circle,fill,inner sep=\thickness pt]{};
    \node at (4,5) [circle,fill,inner sep=\thickness pt]{};
    \node at (5,5) [circle,fill,inner sep=\thickness pt]{};
    \node at (6,5) [circle,fill,inner sep=\thickness pt]{};
    
    \node at (1,6) [circle,fill,inner sep=\thickness pt]{};
    \node at (2,6) [circle,fill,inner sep=\thickness pt]{};
    \node at (3,6) [circle,fill,inner sep=\thickness pt]{};
    \node at (4,6) [circle,fill,inner sep=\thickness pt]{};
    \node at (5,6) [circle,fill,inner sep=\thickness pt]{};
    \node at (6,6) [circle,fill,inner sep=\thickness pt]{};

    \draw[rounded corners=10pt, ultra thick, red] (1-0.32, 1-0.32) rectangle (3+0.32, 3+0.32) {};
    \draw[rounded corners=10pt, ultra thick, red] (4-0.32, 4-0.32) rectangle (6+0.32, 6+0.32) {};
    
    \draw[rounded corners=10pt, ultra thick, blue] (1-0.32, 4-0.32) rectangle (3+0.32, 6+0.32) {};
    \draw[rounded corners=10pt, ultra thick, blue] (4-0.32, 1-0.32) rectangle (6+0.25, 3+0.32) {};
    
    \draw[rounded corners=10pt, ultra thick, green] (0-0.32, 1-0.32) rectangle (0+0.32, 3+0.32) {};
    \draw[rounded corners=10pt, ultra thick, green] (0-0.32, 4-0.32) rectangle (0+0.32, 6+0.32) {};

    \draw[rounded corners=10pt, ultra thick, purple] (1-0.25, 0-0.32) rectangle (3+0.25, 0+0.32) {};
    \draw[rounded corners=10pt, ultra thick, purple] (4-0.25, 0-0.32) rectangle (6+0.25, 0+0.32) {};
    
        \node [color=red] at (2,1.5) {\Large 1a};
        \node [color=blue] at (5,1.5) {\Large 2b};
        \node [color=blue] at (2,4.5) {\Large 2a};
        \node [color=red] at (5,4.5) {\Large 1b};
        
        \node [color=green] at (-0.75,2) {\Large 3a};
        \node [color=green] at (-0.75,5) {\Large 3b};
        
        \node [color=purple] at (2, -0.75) {\Large 4a};
        \node [color=purple] at (5,-0.75) {\Large 4b};
    \end{tikzpicture}
     };
    \node (B) at (6,0) {\includestandalone[width=0.48\textwidth]{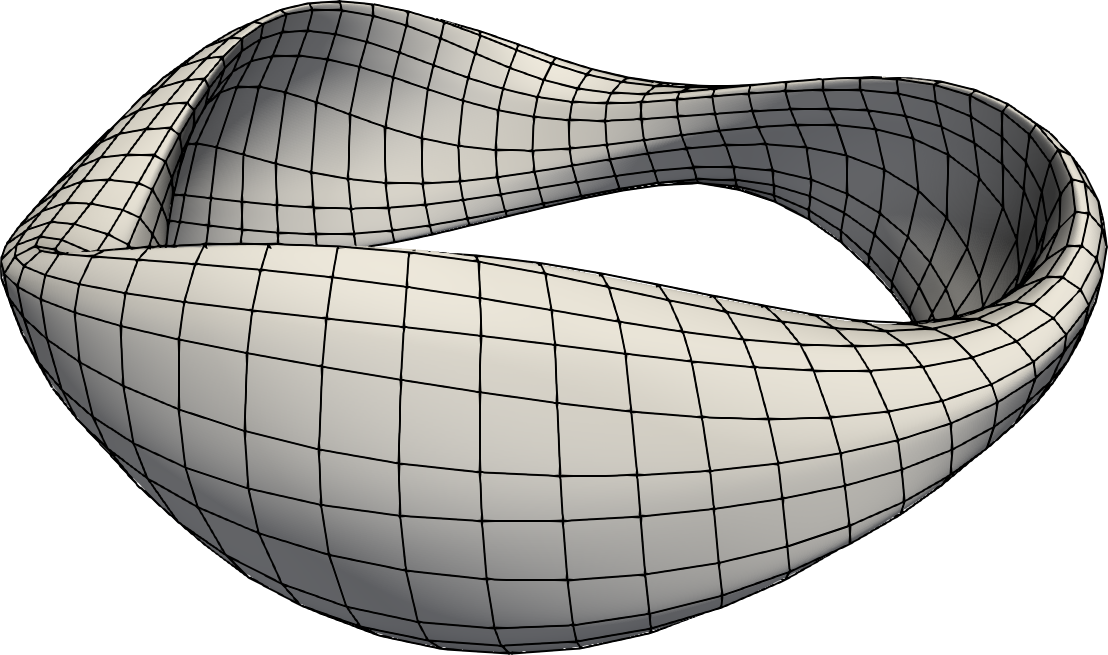} };
    \end{tikzpicture}
    \caption{\emph{Left:} Collocation points on a surface with $n_{\text{tor}} = m_{\text{pol}} = 3$, grouped by color.  When stellarator symmetry is assumed, we choose one green, blue, purple, and red grouping of collocation points.  Choosing points from two groups of the same color results in redundant equations.  The $y$ and $z$ components of the residual at the origin are always included as the $x$ component is zero due to stellarator symmetry.
    \emph{Right:} Lines of constant Boozer angles $\varphi, \theta$ on a magnetic surface of the field generated by a subset of the NCSX coils (section 6 of \cite{Giuliani2020Singlestage}).  
    Arrows indicate increasing $\varphi$ and $\theta$, which respectively follow the long and short way around the torus.
    Note that the toroidal Boozer angle is not equal to the cylindrical toroidal angle.
    }
    \label{fig:isoangles}
\end{figure}
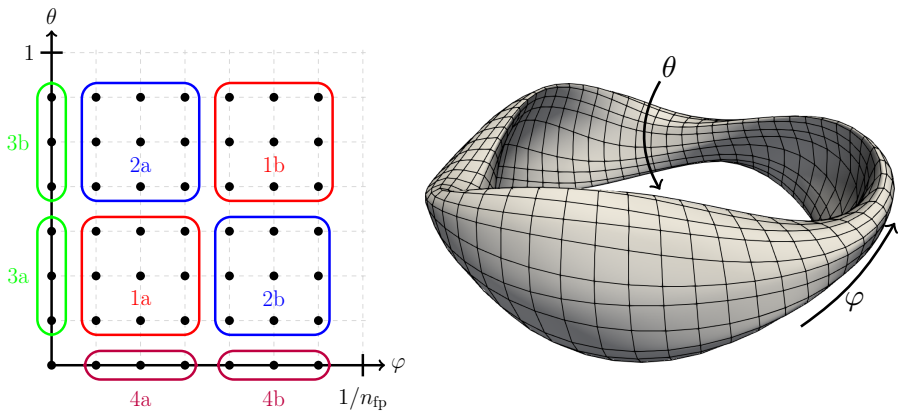

Given a magnetic field with flux surfaces, Boozer coordinates $(\varphi, \theta)$ satisfy
\begin{equation} \label{eq:1}
\mathbf B = \nabla \Psi \times \nabla \theta + \iota \nabla \varphi \times \nabla \Psi,
\end{equation}
where $\Psi$ is the toroidal flux, derivatives are taken with respect to Cartesian coordinates~\citep{d2012flux}, and we assume that $\iota \neq 0$.
Note that \eqref{eq:1} does not apply in regions in which the magnetic field lines are stochastic and fill a volume.
In a vacuum, the magnetic field can be written as
\begin{equation} \label{eq:2}
\mathbf B = G \nabla \varphi,
\end{equation}
where $G$ is a constant.  This is because $\nabla \times \mathbf B=0$ and $\nabla \cdot \mathbf B=0$ implies that $\mathbf B = \nabla V$ for some potential $V = G \varphi$.
Taking the dot product of both sides of \eqref{eq:1} and \eqref{eq:2} with each other and dividing by $G$, we obtain
\begin{equation}\label{eq:3}
    \nabla \Psi \cdot \nabla \theta \times \nabla \varphi = \frac{B^2}{G},
\end{equation}
where the field strength is given by $B := \|\mathbf{B}\|$.  Using \eqref{eq:3} with the following dual relations~\citep{RevModPhys.76.1071},
$$
\nabla \varphi \times \nabla \Psi \frac{G}{B^2}= \frac{\partial \bm \Sigma}{\partial \theta}, ~~ \nabla \Psi \times \nabla \theta \frac{G}{B^2}= \frac{\partial \bm \Sigma} {\partial \varphi}, ~ \text{ and } \nabla \theta \times \nabla \varphi \frac{G}{B^2}= \frac{\partial \bm \Sigma}{\partial \Psi}, 
$$
we conclude that surfaces represented in Boozer angles must satisfy
\begin{subequations}\label{eq:cons}
\begin{align} 
G\mathbf{B} -  \|\mathbf{B}\|^2\left(\frac{\partial \bm \Sigma}{\partial \varphi}+\iota \frac{\partial \bm \Sigma}{\partial \theta} \right) &= 0, \label{eq:cons1} \\
V(\bm \Sigma)- V_{\text{target}}  &= 0. \label{eq:cons2}
\end{align}
\end{subequations}
 Here, $\bm B$ is the magnetic field on $\bm\Sigma$.
 In \eqref{eq:cons2}, $V(\bm\Sigma)$ is the volume enclosed by $\bm\Sigma$ and $V_{\text{target}}\in \mathbb R \backslash \{0\}$ is a given target volume. The label constraint \eqref{eq:cons2} is necessary to make \eqref{eq:cons} a closed system. 
 For the volume, we use the formula
\begin{equation}
    V(\bm \Sigma) := \frac{1}{3}\int_0^1 \int_0^1  \bm \Sigma \cdot \left(\frac{\partial \bm \Sigma}{\partial \varphi} \times \frac{\partial \bm \Sigma}{\partial \theta} \right) ~d\varphi ~d\theta.
\end{equation}
The unknowns in \eqref{eq:cons} are $G\in \mathbb R$, the surface $\bm \Sigma$, and the rotational transform $\iota$.  The solution to \eqref{eq:cons} is a magnetic surface for the field $\mathbf B$ that is parametrized in Boozer angles.
In principle, we know that $G = n_{\text{fp}}(\mu_0/\pi) \sum^{N_c}_{k=1}I_k $ for an exact stellarator symmetric magnetic surface, where $I_k$ is the current in the $k$th coil. 
However, we include $G$ as an unknown since this may not be true for finite dimensional approximations of \eqref{eq:cons}.
As illustration of our parameterization, we show a surface that is everywhere tangent to a magnetic field generated by a subset of the NCSX coils in Figure \ref{fig:isoangles}, as well as curves of constant Boozer angle $\varphi, \theta$ on that surface.
The Boozer angles $\varphi, \theta$ increase, respectively, the long and short way around the toroidal surface.

In the next sections, we discuss the nonlinear system resulting from replacing $\bm \Sigma$ with the finite-dimensional surface $\bm\Sigma_s(\varphi,\theta)$ from section \ref{sec:surface-param} and enforcing \eqref{eq:cons} on a grid of collocation points. Depending on the number of degrees of freedom in the parametrization and the number of collocation points, we either obtain a balanced system where the number of degrees of freedom is the same as the number of nonlinear equations, or an overdetermined system where the number of equations is larger than the number of unknowns.
The derived PDE holds only in regions of the magnetic field where Boozer coordinates can be constructed.  Despite this, we show an approach to nevertheless compute surfaces in regions of the magnetic field where nested flux surfaces do not exist.

\subsection{Balanced system}\label{sec:BoozerExact}
We obtain a balanced system by using a parametrization $\bm\Sigma_s$ of $\bm \Sigma$ with unknowns $\mathbf s\in \mathbb R^{n_s}$, requiring \eqref{eq:cons} to be satisfied at the same number of collocation points, i.e.,
\begin{equation}
\begin{aligned} \label{eq:surface_solve1}
\mathbf{r}_{i,j}(\mathbf s) := G\mathbf{B}_{i,j} -\|\mathbf{B}_{i,j}\|^2\left(\frac{\partial \bm \Sigma_{s,i,j}}{\partial \varphi}+\iota  \frac{\partial \bm \Sigma_{s,i,j}}{\partial \theta} \right) &= 0, ~\forall (i,j) \in \mathcal{C} \\
V(\bm\Sigma_s)- V_{\text{target}}  &= 0,
\end{aligned}
\end{equation}
where $\mathcal{C}$ is the set of indices of collocation points on a regular grid, $\bm\Sigma_{s,i,j} = \bm\Sigma_{s}(\varphi_{i,j}, \theta_{i,j})$, and $V(\bm\Sigma_s)$ is the volume enclosed by the surface, computed as
\begin{equation} 
    V(\bm\Sigma_s) = \frac{1}{3 n'_{\varphi} n'_{\theta}} \sum_i \sum_j   \bm \Sigma_{s,i,j} \cdot \left(\frac{\partial \bm \Sigma_{s,i,j}}{\partial \varphi} \times \frac{\partial \bm \Sigma_{s,i,j}}{\partial \theta} \right) \text{ for } (i,j) \in \mathcal{C}',
\end{equation}
where $\mathcal{C}'$ is a set of quadrature points on a regular $n'_{\varphi} \times n'_{\theta}$ grid that is possibly different from the ones used by the collocation method $\mathcal{C}$ to ensure an accurate computation of the volume enclosed by the surface.
The equations \eqref{eq:surface_solve1} require that the residual at a set of collocation points $(\varphi_{i,j}, \theta_{i,j})$ be zero.  
We place a grid of $(2n_{\text{tor}} +1) \times (2m_{\text{pol}}+1)$ equispaced collocation points on $[0,1/n_{\text{nfp}})\times[0,1)$ and refer to surfaces that satisfy \eqref{eq:surface_solve1} at each point on that grid as BoozerExact surfaces.

Assuming stellarator symmetry, a number of these points are redundant and we choose a subset of them. 
First, the $x$-component of the residual at the quadrature point associated to the origin, $i = j = 0$, or  $\varphi = \theta = 0$, is always zero for all surfaces, so this equation can be ignored.  
To see why this is the case, note that the $x$-component of stellarator symmetric magnetic fields is zero when $\varphi=\theta = 0$.  Similarly, differentiating the first equation in \eqref{eq:ss} with respect to $\varphi$ or $\theta$ shows that the $x$-component of the surface tangents are zero when $\varphi=\theta = 0$.
Second, if we include $(\varphi_{i,j}, \theta_{i,j})$ then the collocation point associated to $(-\varphi_{i,j}, -\theta_{i,j})$ is unnecessary.  
This is because the residual evaluated at $(\varphi_{i,j}, \theta_{i,j})$ is the residual at $(-\varphi_{i,j}, -\theta_{i,j})$ subject to a reflection.
This is illustrated on the left in Figure \ref{fig:isoangles}, where the collocation points are grouped by color.  We are free to choose one green, blue, purple, and red grouping.  Choosing points from two groups of the same color would result in redundant equations.  In all examples, we place collocation points on $[0,1/2n_{\text{fp}})\times[0,1)$ from zones 1a, 2a, 3a, and 4a, resulting in, respectively, $n_{\varphi} = n_{\mathrm{tor}} + 1$ and $m_{\theta} = 2 m_{\mathrm{pol}} + 1$ quadrature points in the toroidal and poloidal directions (before removing redundant quadrature points in group 3b and at the origin).

After removing the redundant collocation points, the number of equations that must be satisfied is
\begin{equation}
    3[2n_{\text{tor}}m_{\text{pol}} + m_{\text{pol}} + n_{\text{tor}} + 1] - 1 + 1 = 3[2n_{\text{tor}}m_{\text{pol}} + m_{\text{pol}} + n_{\text{tor}} + 1].
\end{equation}
This is the same as $n_s+2$, the number of geometric degrees of freedom in the parametrization $n_s$ with $\iota$ and $G$. Since this is a balanced system of nonlinear equations, we can use Newton's method to solve it.

\subsection{Overdetermined system}\label{sec:BoozerLS}
As we will show in numerical examples, the above balanced approach can fail to converge when the magnetic field lines have chaotic regions and islands. For such problems, we propose an alternative approach in which
we relax the requirement that the residual be zero everywhere. Instead, we compute a surface that solves \eqref{eq:cons} in a least-squares sense.  That is, it minimizes
\begin{equation} \label{eq:boozerls}
    f(\mathbf{s}) := \frac{1}{2}   \sum_{i,j} \|\bm r_{i,j}(\bm s) \|^2 +  \frac{1}{2} w \left( V(\bm\Sigma_s)- V_{\text{target}} \right)^2.
\end{equation}
Here, the nonlinear least squares objective contains the sum of squares of \eqref{eq:cons} at a number of collocation points with a penalty term with weight $w>0$ to approximately enforce the surface volume condition \eqref{eq:cons2}.  The first-order optimality condition for a minimizer of $f$ is
\begin{equation}\label{eq:surface_solve2}
\nabla_\mathbf{s} f = 0
\end{equation}
and the optimization problem can be solved using a Levenberg-Marquardt or Newton's method.  In numerical experiments, we typically first compute a solution with the Levenberg-Marquardt method, which is known to work robustly for nonlinear least squares problems \citep{nocedal2006numerical}.  If the stationarity condition \eqref{eq:surface_solve2} is not satisfied to sufficient precision, then we further improve the solution using a few steps of Newton's method.
We refer to surfaces that satisfy \eqref{eq:surface_solve2} as BoozerLS (``Boozer Least Squares'') surfaces.
Note that the BoozerLS surface has similarities with the quadratic flux
    minimizing (QFM) surfaces of~\cite{dewar2010unified}. QFM surfaces are
    obtained by finding surfaces so that $(\bm B\cdot\bm n)^2$ is minimized on
    them.  This results in a surfaces that are approximately tangential to the magnetic field. 
    Surfaces that satisfy \eqref{eq:cons1} have the stricter
    requirement that the magnetic field
    must be aligned in the direction $\frac{\partial \bm
    \Sigma}{\partial \varphi}+\iota \frac{\partial \bm \Sigma}{\partial \theta}$ and not just any tangential direction.
In all examples, we place equispaced collocation points on $[0,1/2n_{\text{fp}})\times[0,1)$, resulting in respectively $n_{\varphi} = n_{\mathrm{tor}} + 6$ and $m_{\theta} = 2 m_{\mathrm{pol}} + 6$ quadrature points in the toroidal and poloidal directions.

\subsection{Example: Surfaces of an NCSX-like magnetic field}\label{sec:surface_example}
In this section, we use a challenging magnetic field to study the behaviour of the BoozerExact and BoozerLS approaches for computing surfaces. 
Specifically, we use the 18 nonplanar coils and currents, corresponding to the C09R00 boundary configuration, of the national compact stellarator experiment (NCSX) to generate a vacuum magnetic field.  
The original NCSX design included planar toroidal and poloidal field coils that we do not use here.  We have also previously considered this set up in section 6 of \cite{Giuliani2020Singlestage}.
The field is challenging as it does not have nested flux surfaces everywhere, and presents an island chain in the neighborhood of the low-order rational $\iota=3/7$.
The magnetic field $\mathbf B$ on the surface $\bm \Sigma$ in \eqref{eq:cons1} is evaluated using the Biot-Savart law since we are working with vacuum fields.
Newton's method requires the Jacobian of \eqref{eq:surface_solve1} or \eqref{eq:surface_solve2}, which are functions of the magnetic field and its first spatial derivatives for \eqref{eq:surface_solve1}, and of its first and second spatial derivatives for \eqref{eq:surface_solve2}.  Since the field is computed with the Biot-Savart law, we have explicit formulas to compute these quantities.  

Due to the local convergence properties of the Newton and Levenberg-Marquardt algorithms to compute surfaces, they require initializations that should not be too far from the target surface. We use a continuation approach, i.e., we first compute a surface close to the axis, i.e., with a small value of $V_\text{target}$.
For that purpose, we generate a toroidal surface centered on the magnetic axis with a fixed minor radius. This does not correspond to a magnetic flux surface. However, in the neighborhood of the magnetic axis it is typically close enough for Newton's method to converge. We then use previously computed surfaces as initialization for finding the surface for the next larger $V_\text{target}$.  

\begin{figure}
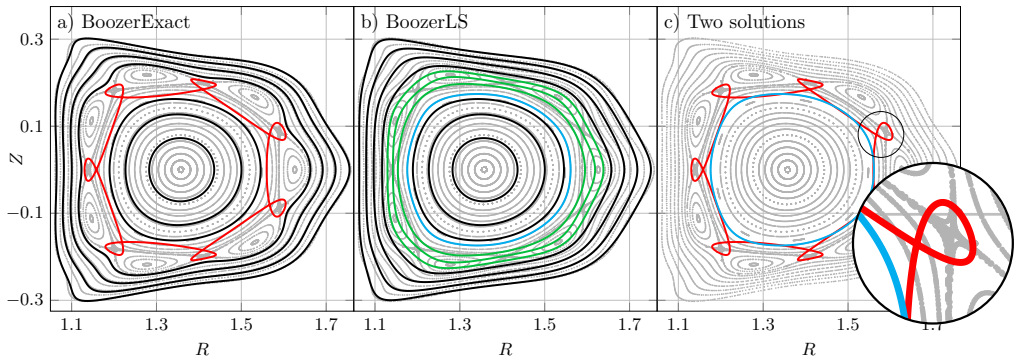

    \centering
\includestandalone[width=\textwidth]{computing_surfaces/pb1/combined_spy}
    \caption{Poincar\'e plots (in grey) and surface cross sections for
      the field generated from a set of NCSX modular coils.
      at $\phi = \pi/3$.  In a), we
      show cross sections of surfaces computed with the BoozerExact
      approach (in red and black).  The red cross section is
      self-intersecting as it is close to an island chain. In b) we
      show cross section obtained with the BoozerLS algorithm (in
      black, green and cyan), which results
      in nested surfaces that do not self-intersect in the presence of
      islands. The red and cyan cross sections in a) and b) have
      the same target surface label, but are computed using
      BoozerExact and BoozerLS formulations, respectively.  Green
      cross sections correspond to additional surfaces not shown in
      a).  In c), we show two BoozerExact surfaces that have the same
      enclosed surface volume.  }
    \label{fig:boozerExactvsBoozerLS}
\end{figure}

In Figure~\ref{fig:boozerExactvsBoozerLS}a, we show cross sections of surfaces obtained with the BoozerExact approach, i.e., using the same number of unknowns as equations. In particular, we solve \eqref{eq:surface_solve1} with $m_{\text{pol}} = n_{\text{tor}} = 11$. As can be seen, when the Poincar\'e plot suggests the existence of magnetic surfaces, the BoozerExact approach converges to surfaces that align well with the Poincar\'e plot.  In chaotic regions or regions containing magnetic islands, the approach may not converge, or it may find self-intersecting surfaces (as shown in red in the figure).

We next show that the BoozerLS approach is more robust in the presence of islands and chaotic regions. We again use $m_{\text{pol}} = n_{\text{tor}} = 11$ and the volume constraint penalty $w=1000$, which results in a small relative error in the volume contained by the surface, ranging from $10^{-9}$ to $10^{-5}$. The resulting cross sections are shown in Figure \ref{fig:boozerExactvsBoozerLS}b.
It can be seen that when the Poincar\'e plot indicates the existence of good surfaces, the BoozerLS surface cross sections visually coincide with the Poincar\'e plot and with the cross sections obtained with the balanced BoozerExact approach. In other regions, the BoozerLS approach results in layered smooth surfaces that do not have self-intersections (green curves). 
BoozerLS surfaces can either be computed using a standard continuation procedure, or alternatively, using possibly self-intersecting BoozerExact surfaces as initial guesses.

Finally, we numerically show that BoozerExact surfaces, i.e., solutions to \eqref{eq:surface_solve1}, are not necessarily unique in the presence of chaos.  This is more of theoretical interest than a fundamental problem as this is still a practical approach for computing magnetic surfaces when they exist.  In addition, we will show in section \ref{sec:qs} that it also can be used to optimize coils for quasi-symmetry. 
In Figure \ref{fig:boozerExactvsBoozerLS}c, we show two BoozerExact surface cross sections with the same target volume enclosed, $V_{\text{target}}=1$.  The red surfaces in Figure~\ref{fig:boozerExactvsBoozerLS}a and c are the same.  The cyan cross section in Figure~\ref{fig:boozerExactvsBoozerLS}c is a BoozerExact surface initialized from the cyan BoozerLS surface in Figure~\ref{fig:boozerExactvsBoozerLS}b. 
It is still possible that the BoozerLS solver does not converge or for a self-intersecting BoozerLS surface to be found. However we find that this happens less frequently than it does for BoozerExact surfaces.

\begin{figure}
    \centering
\begin{tikzpicture}[spy using outlines={circle, magnification=3.5}]
    \node (11) at (0,0) {\pgfplotsset{every axis title/.style={below right,at={(0,1)}}}
\newcommand{\vasymptote}[2][]{
    \draw [densely dashed,#1] ({rel axis cs:0,0} -| {axis cs:#2,0}) -- ({rel axis cs:0,1} -| {axis cs:#2,0});
}
\begin{tikzpicture}
    \begin{groupplot}[
            group style={
                group name=my plots,
                group size=2 by 1,
                xlabels at=edge bottom,
                ylabels at=edge left,
                horizontal sep=2cm,vertical sep=0cm,},
            xlabel={Volume},
            grid=both,
            label style={font=\Large},
            legend style={at={(2.35,0.5)},anchor=west}
        ]
        \nextgroupplot[
            ylabel={2-norm of residual/$\sqrt{n_\phi n_\theta}$},
            ymode=log
        ]%
        
        \pgfplotsset{every tick label/.append style={font=\Large}}
        \addplot[line width=1.5pt, color=Dark2-A] table[x index=0, y expr=\thisrowno{1}/sqrt{176}] {computing_surfaces/pb2/data/res_5.txt};
        \addplot[line width=1.5pt, color=Dark2-B] table[x index=0, y expr=\thisrowno{1}/sqrt{416}] {computing_surfaces/pb2/data/res_10.txt};
        \addplot[line width=1.5pt, color=Dark2-C] table[x index=0, y expr=\thisrowno{1}/sqrt{756}] {computing_surfaces/pb2/data/res_15.txt};
        \addplot[line width=1.5pt, color=Dark2-D] table[x index=0, y expr=\thisrowno{1}/sqrt{1196}] {computing_surfaces/pb2/data/res_20.txt};
        \addplot[line width=1.5pt, color=Dark2-E] table[x index=0, y expr=\thisrowno{1}/sqrt{1736}] {computing_surfaces/pb2/data/res_25.txt};
        \vasymptote{1.529};
        
        \addlegendentry{$n_{\mathrm{tor}} = m_{\mathrm{pol}} = 5$};
        \addlegendentry{$n_{\mathrm{tor}} = m_{\mathrm{pol}} = 10$};
        \addlegendentry{$n_{\mathrm{tor}} = m_{\mathrm{pol}} = 15$};
        \addlegendentry{$n_{\mathrm{tor}} = m_{\mathrm{pol}} = 20$};
        \addlegendentry{$n_{\mathrm{tor}} = m_{\mathrm{pol}} = 25$};

        \nextgroupplot[
            ylabel={$\iota$},
            xticklabel pos=bottom
        ]%
        \pgfplotsset{every tick label/.append style={font=\Large}}
        \addplot[line width=1.5pt, color=Dark2-A] table[x index=0, y index=1] {computing_surfaces/pb2/data/i_5.txt};
        \addplot[line width=1.5pt, color=Dark2-B] table[x index=0, y index=1] {computing_surfaces/pb2/data/i_10.txt};
        \addplot[line width=1.5pt, color=Dark2-C] table[x index=0, y index=1] {computing_surfaces/pb2/data/i_15.txt};
        \addplot[line width=1.5pt, color=Dark2-D] table[x index=0, y index=1] {computing_surfaces/pb2/data/i_20.txt};
        \addplot[line width=1.5pt, color=Dark2-E] table[x index=0, y index=1] {computing_surfaces/pb2/data/i_25.txt};
        \draw[densely dashed]  (0,3/7) -- (1.529, 3/7); 
        \draw[densely dashed]  (1.529,3/7) -- (1.529,0); 
    \end{groupplot}
\node at (-1.25cm, -0.75)(a) {};
\node at (-1.25cm, 1)(a) {};
\end{tikzpicture} }; 
\end{tikzpicture}
    \caption{BoozerLS residual in the NCSX-lite configuration with increasing number of Fourier modes used to represent the surface.  
    First figure shows the BoozerLS residuals as a function of the surface label. We observe that it stagnates in the neighborhood of $\iota=3/7$, which corresponds to an island chain.
    Second figure shows the rotational transform with respect to surface label for different resolutions.  
    The vertical dashed line approximately indicates the volume for which $\iota=3/7$ is achieved.
    }
    \label{fig:boozerLSresidual}
\end{figure}
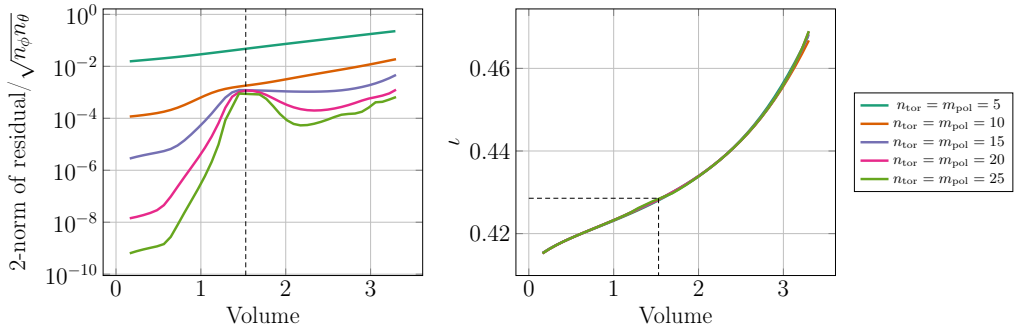

To further study the BoozerLS method in the presence of islands or chaotic fields, we examine the behavior of the residual as the number of Fourier modes ($n_{\text{tor}}$, $m_{\text{pol}}$) and collocation points used in the parametrization of the surface increases (Figure \ref{fig:boozerLSresidual}). 
As the number of degrees of freedom and collocation points in $\bm \Sigma_s$ increase, the residual decreases.  
However, this is only true for surfaces that do not pass through island chains.
In the neighborhood of the island chain, we observe that the residual does not decrease with increasing surface complexity and appears to stagnate.
This hints at an approach to optimize for magnetic fields without islands: adding the square of the residual at the collocation points of the BoozerLS surfaces to the objective function. 
This approach is not used in the coil optimization discussed next, which relies on BoozerExact surfaces. However, it will be the subject of future work.

\section{Optimization for quasi-axisymmetry on surfaces} \label{sec:qs}
To target quasi-symmetry on magnetic surfaces within an optimization problem, we need a measure of quasi-symmetry on a given surface.
For that purpose, we recall that the field strength $B:=\|\mathbf B\|$ of a quasi-axisymmetric field on a magnetic surface in Boozer coordinates only depends on the angle $\theta$. This motivates us to consider a decomposition into a quasi-axisymmetric and non-quasi-axisymmetric component
$$
B(\varphi,\theta) = B_{\text{QS}}(\theta) + B_{\text{non-QS}}(\varphi,\theta).
$$
This decomposition is not unique, and we choose $B_{\text{QS}}(\theta)$ such that it is closest to the field strength $B$ when measured using the $L_2$-norm $(\int_{\bm\Sigma_s} f^2 ~dS)^{1/2}$, that is,
$$
B_{\text{QS}}(\theta) = \frac{\int^{1/n_{\text{fp}}}_{0} B(\bm\Sigma_s(\varphi,\theta))~\| 
\frac{\partial \bm\Sigma_s}{\partial \varphi} \times \frac{\partial \bm\Sigma_s}{\partial \theta} \|~d\varphi}{\int^{1/n_{\text{fp}}}_{0} ~ \| 
\frac{\partial \bm\Sigma_s}{\partial \varphi} \times \frac{\partial \bm\Sigma_s}{\partial \theta} \|~d\varphi}.
$$
Evaluating these quantities is straightforward because our surfaces are already parametrized in Boozer coordinates.
\subsection{Optimization problem formulation}
\label{sec:opt}
Now that we are able to compute surfaces and a measure for quasi-symmetry, the goal is to formulate and solve optimization problems for coils generating magnetic fields with good quasi-symmetry properties on a large number of toroidal surfaces.
We seek a design with $N_c$ independent modular coils, to which stellarator and rotational symmetries are applied.  After application of these symmetries, the stellarator is made up of $2n_{\mathrm{fp}}N_c$ modular coils. Each coil is parametrized with a current and $n_c$ geometric degrees of freedom using a Fourier basis as in \cite{Giuliani2020Singlestage}. All coil degrees of freedom and their currents are summarized into a vector $\mathbf c\in \mathbb R^{N_cn_c + N_c}$. Our guiding principle for designing the coils is to target quasi-symmetry of the magnetic field induced by these coils on $N_s$ surfaces. Each surface uses the parametrization from Section \ref{sec:surface-param} with $n_s$ geometric degrees of freedom. The resulting surface Fourier coefficients, $\iota$, and $G$ are summarized into a vector $\mathbf s\in \mathbb R^{N_sn_s+2N_s}$. In the optimization formulation below, we use BoozerExact surfaces (Section \ref{sec:BoozerExact}). 
Finally, the optimization also includes various 
(inequality) constraints for the coils that are motivated by engineering considerations, which are critical in terms of the feasibility and cost-efficiency of a design \citep{Strykowsky09,Neilson10,Klinger2013}. These coil constraints are typically not difficult to compute as they explicitly depend on the coil degrees of freedom.  This is in contrast to the terms that depend on the surfaces, which  have an implicit dependence on the coil degrees of freedom.  The optimization problem is
\begin{subequations}\label{eq:optprob}
\begin{align}
    \min_{\mathbf c \in \mathbb R^{N_{\!c} n_{c} + N_c}, \: \mathbf s\in \mathbb R^{N_{\!s} n_s + 2 N_s}} & ~ \hat J (\mathbf c, \mathbf s) \label{eq:optprob:J}\\
    \text{subject to }  \eqref{eq:surface_solve1} &\text{ on each of the $N_s$ surfaces,}  \label{eq:optprob:2}\\
    \frac{1}{N_s}\sum_{k=1}^{N_s}\iota_k &= \overline{\iota}, \label{eq:optprob:3}\\
    \Psi &= \Psi_0, \label{eq:optprob:4}\\
    R_{\text{major}} &= R_0, \label{eq:optprob:5}\\
    \sum_{i = 1}^{N_c} L^{(i)}_{c} &\leq L_{\max}, \label{eq:optprob:6}\\
    \kappa_i &\leq \kappa_{\max}, ~ i = 1, \ldots, N_c, \label{eq:optprob:7}\\
    \frac{1}{L^{(i)}_{c}}\int_{\bm\Gamma^{(i)}} \kappa_i^2 ~dl &\leq \kappa_{\mathrm{msc}}, ~ i = 1, \ldots, N_c, \label{eq:optprob:8}\\
    \| \bm\Gamma^{(i)}- \bm\Gamma^{(j)} \| &\geq d_{\min} ~ \text{ for } i \neq j,\label{eq:optprob:9}
\end{align}
\end{subequations}
where the objective function is the average (normalized) non quasi-axisymmetry on the surfaces $\bm \Sigma_{s,k} ~ k= 1, \hdots, N_s$:
\begin{align}
\begin{split}
\hat{J}(\mathbf{c},\mathbf{s}) &= \frac{1}{N_s}\sum^{N_s}_{k=1} \frac{\int_{\bm\Sigma_{s,k} }B_{\text{non-QS}, k}(\mathbf{c}, \mathbf{s})^2~dS }{ \int_{\bm\Sigma_{s,k} }B_{\text{QS}, k}(\mathbf{c}, \mathbf{s})^2~dS}.
\end{split}\label{eq:objf4}
\end{align}
The above objective is the average ratio of the squared 2-norm of the field magnitude's non-quasisymmetric and quasisymmetric components.
The objective is subject to the surface constraints \eqref{eq:optprob:2}--\eqref{eq:optprob:5} and the coil constraints \eqref{eq:optprob:6}--\eqref{eq:optprob:9}. In particular, \eqref{eq:optprob:3} ensures that the average rotational transform across the surfaces is $\overline{\iota}$. \eqref{eq:optprob:4} fixes the toroidal flux on the innermost surface to a given value $\Psi_0$ and prevents the currents from going to zero.  \eqref{eq:optprob:5} fixes the major radius on the innermost surface to a given $R_0$ and prevents the length scale of the stellarator from changing.  \eqref{eq:optprob:6} prevents the sum of the independent modular coil lengths  $\sum_{i = 1}^{N_c} L^{(i)}_{c}(\mathbf{c})$ from exceeding a given value $L_{\max}>0$.  \eqref{eq:optprob:7} and \eqref{eq:optprob:8}, respectively, prevent the curvature and mean squared curvature on each coil from exceeding values $\kappa_{\max}$ and $\kappa_{\mathrm{msc}}$.  Finally, \eqref{eq:optprob:9} ensures that the coils stay at least $d_{\min}>0$ away from one another.

\subsection{Computational aspects}
Our computational approach to finding minimizers for \eqref{eq:optprob} uses a gradient-based descent algorithm to minimize the objective. This requires appropriate gradients of the objective, which take into account the equality PDE-constraints \eqref{eq:optprob:2} describing the surfaces, as well as the remaining equality and inequality constraints \eqref{eq:optprob:3}--\eqref{eq:optprob:9}.
To enforce the PDE-constraints \eqref{eq:optprob:2}, we use the method of Lagrange multipliers \citep{Troltzsch10, Giuliani2020Singlestage}. For that purpose, we consider the objective as a function just of the coil degrees of freedom, i.e.,
$J(\mathbf{c}) := \hat J(\mathbf{c}, \mathbf{s}(\mathbf{c}))$ by considering the surfaces as functions of the coils, i.e., $\mathbf s = \mathbf s(\mathbf c)$ implicitly defined through the solution of \eqref{eq:surface_solve1}.
Computing the gradient of $J$ with respect to $\mathbf c$ can be done efficiently by introducing Lagrange multipliers and using the adjoint method. Computing gradients then requires solving two systems of equations per surface, both with $n_{s}+2$ equations: the nonlinear system of equations \eqref{eq:surface_solve1} and a linear system for the adjoint variable. This method compares favorably to finite difference approaches for computing gradients, where the number of times \eqref{eq:surface_solve1} must be solved scales with $n_c$, the number of degrees of freedom used to represent the coils.

The constraints \eqref{eq:optprob:3}--\eqref{eq:optprob:9} are enforced approximately using a penalty method, with penalties that only become active when the constraint is violated. 
We next detail the penalty terms used, and discuss a practical strategy to choose and adjust the weights for these penalties at the end of this section.  

For the penalty on the sum of the independent modular coil lengths, we use the standard quadratic penalty
\begin{equation*}
    \frac{1}{2}\left [\max \left( 0, \sum_{i = 1}^{N_c} L^{(i)}_{c}(\mathbf{c}) - L_{\max} \right) \right]^2.
\end{equation*}
The penalty on the maximum curvature for coil $\bm \Gamma^{(i)}$ is
\begin{equation*}
    \sum_{i=1}^{N_c}\frac{1}{2} \int\limits_{\bm \Gamma^{(i)}} \left [\max \left( 0, \kappa_i(\mathbf c) - \kappa_{\text{max}} \right) \right] ^2~dl_i,
\end{equation*}
and, as in \cite{Giuliani2020Singlestage}, the penalty on the minimum pairwise coil distance is
\begin{equation}
\frac{1} 2 
    \sum_{i\neq j} \int\limits_{\bm \Gamma^{(i)}}\int\limits_{\bm \Gamma^{(j)}} \left [ \max(0, d_\mathrm{min} - \| \bm \Gamma^{(i)}-\bm  \Gamma^{(j)} \|_2) \right]^2 ~ dl_j\, dl_i. \label{eqn:min_dist}
\end{equation}
For the equality constraints, we add quadratic penalties to the objective. For instance, to prevent the currents in the coils from becoming zero, for  
the toroidal flux $\Psi$ of the innermost surface we use
$$
 \frac{1}{2}\left( \Psi(\mathbf c, \mathbf s)-\Psi_0 \right)^2.
$$
Here, the toroidal flux through a toroidal cross section of constant Boozer angle $\varphi=0$, $S_{\varphi=0}$, of a surface is given by
\begin{equation}
\Psi(\mathbf c, \mathbf s) = \int_{S_{\varphi=0}} \mathbf{B} \cdot \mathbf n ~dS
= \int_0^1 \mathbf{A} \cdot \frac{\partial}{\partial \theta} \bm\Sigma_s(0,\theta) ~d\theta,
\label{eq:torflux}
\end{equation}
where $\mathbf A$ is the vector potential, such that $\mathbf B = \nabla \times \mathbf{A}$, and where we have used Stokes' theorem to turn the surface integral for the flux into a line integral.
Additionally, we include a penalty term on the major radius $R_\mathrm{major}$ of the innermost surface: 
$$
\frac{1}{2}  \left(R_{\mathrm{major}}(\mathbf s)-R_0\right)^2.
$$
This provides a characteristic length scale to the stellarator configurations we optimize, and maintains the aspect ratio of the surfaces in the volume.
Details of how we compute this quantity are given in Appendix \ref{Appen:Major}.  
To constrain the mean rotational transform of the surfaces to be close to a given value $\overline \iota$, we use the penalty term
$$
\frac{1}{2}\left(\frac{1}{N_s}\sum_{k=1}^{N_s}\iota_k(\mathbf s) - \overline{\iota}\right)^2,
$$
where $\iota_k$ is the rotational transform on the $k$th surface.
While here we target the average rotational transform, other choices are possible in our framework, such as constraining any number of surface rotational transforms to a given value or interval. 
Finally, we include a penalty that favors coils with an approximately uniform incremental arclength
$$
\sum^{N_c}_{i=1}\int^1_0 (\|\bm \Gamma^{(i)'}\|_2 - t^{(i)})^2~d\xi,
$$
with $t^{(i)} = \int^1_0 \|\bm \Gamma^{(i)'}\|_2~d\xi$,  
where $\xi\in [0,1)$ is the angle parameterization of the coil.
This prevents the incremental arclengths from getting close to zero,
where the length penalty term is nondifferentiable.
Note that it might not be possible to attain a uniform
incremental arclength at all points on the coil. However, this is not
necessary as the purpose of this penalty is
to avoid the above issue.
Numerical
experiments confirm that this term does not substantially affect the
value of the objective over a wide range of penalty weights.

We compute analytical gradients of the (reduced) objective $J(\mathbf c)$, and use the BFGS method \citep{nocedal2006numerical} with linesearch to find a minimizer. Compared to steepest descent, BFGS has the advantage that it approximates second-order information of the objective, which typically results in faster convergence or, in some cases, avoids that the iteration progress stalls. 

To choose appropriate penalty weights, we follow a simple strategy that appropriately increases these weights throughout the iteration. Namely, if any of the constraints are not satisfied to within a $0.1\%$ relative tolerance after 3,000 iterations of BFGS, we increase the weights on the associated penalty terms by a factor of 10. Additionally, we allow for a maximum number of 45,000 BFGS iterations.
However, since we choose a judicious set of initial weights informed from previous runs, weights are increased rarely.

\subsection{Coil optimization for precise quasi-axisymmetry on surfaces}\label{sec:opt-results}

In this example, we revisit a coil set obtained in
\citep{wechsung2022precise} with the second stage of the standard
two-stage optimization method, where the goal was to produce the
precise QA magnetic field presented in \cite{LandremanPrecise} using a
realistic set of coils. In the first stage, that field was optimized
for precise QA on a toroidal volume with aspect ratio 6, and the
average rotational transform in the volume was 0.42.  The stage II
optimization found $N_c=4$ coils with a minimum pairwise coil distance of
$d_{\min} = \SI{0.1}{\meter}$, and maximum curvature and mean squared curvature of
$\kappa_{\max} = \SI{5}{\meter}^{-1}$ and $\kappa_{\mathrm{msc}} =
\SI{5}{\meter}^{-2}$. Various coil designs are found where the
the sum of the lengths of the 4 base coils is constrained to be below $L_{\max} =
\SI{18}{\meter}, \SI{20}{\meter}, \SI{22}{\meter}, \SI{24}{\meter}$.
Here, the major radius of the device is approximately
\SI{1}{\meter}. In the following optimization, we use the same
values for $d_{\min}$, $\kappa_{\max}$, $\kappa_{\mathrm{msc}}$ as in the 
stage II optimization from \cite{wechsung2022precise}.

Building on the results from this two-stage coil design, we aim at improving the quasi-symmetry on multiple surfaces uniformly distributed throughout the volume using the optimization approach presented in the previous section. 
Since the fields resulting from the coils in \cite{wechsung2022precise} appear to avoid islands, we use BoozerExact (rather than BoozerLS) surfaces in this optimization. 
Should islands appear in the vicinity of the surfaces over the course of the optimization, the BoozerExact surfaces might begin to self-intersect and no longer correspond to flux surfaces of the magnetic field that we are trying to optimize.  However, as will be shown, this does not happen in the final coil designs.
Generalizing the formulation \eqref{eq:optprob} to BoozerLS surfaces, in order to handle magnetic fields with islands in a robust manner, and to optimize the coils to eliminate such islands, will be part of future work.

\begin{figure}
    \centering
    \includegraphics[width=0.65\textwidth]{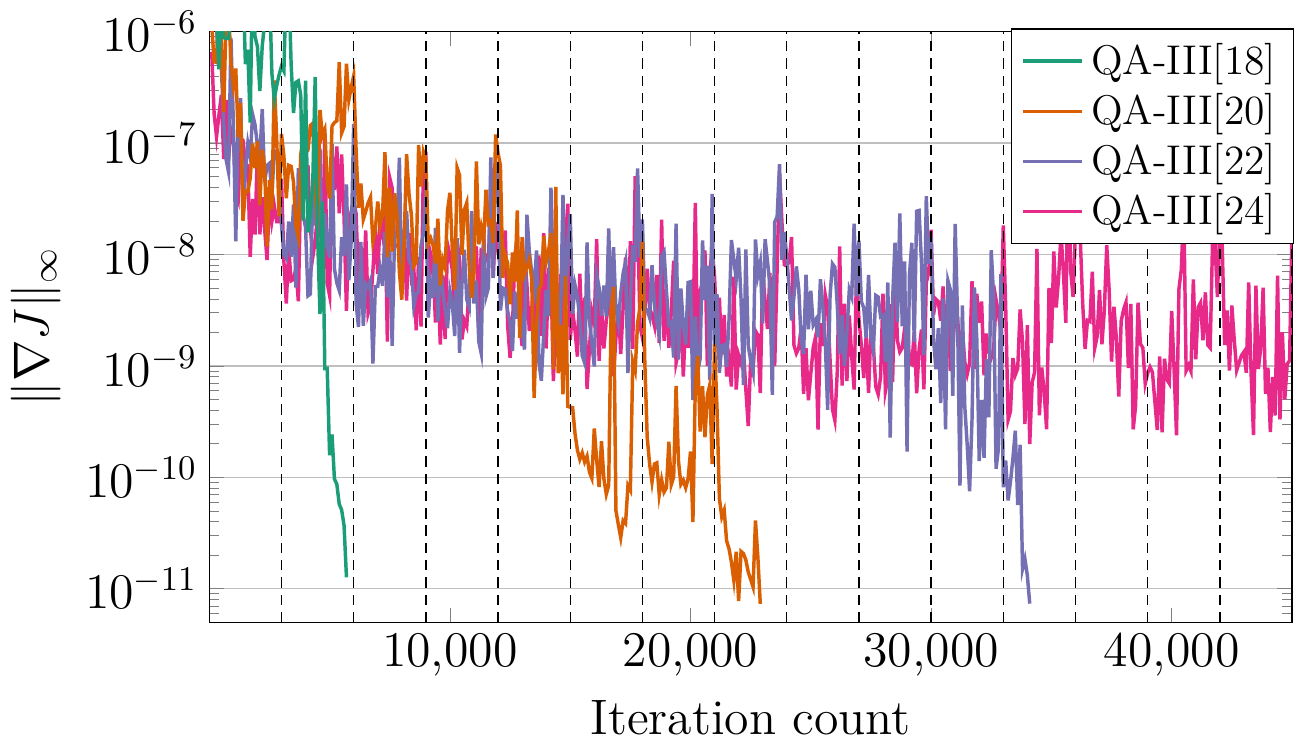}
    \caption{Norm of gradient in each iteration where five surfaces are used.
    Gradients at the initial configuration were $\mathcal{O}(10^{-4})$, but we truncate the $y$-axis in order to better observe convergence to optimality.  Every 3,000 iterations (indicated by dotted vertical lines), the penalty weights are increased until the (in)equality constraints in \eqref{eq:optprob} are satisfied to sufficient precision.
    Since a suitable set of initial weights were chosen, this does not occur frequently.
    }
    \label{fig:five_gradient}
\end{figure}

\newcommand{\ds}{\raisebox{2pt}{\tikz{\draw[black,dashed,line width=0.9pt](0,0) -- (5mm,0);}}}
\newcommand{\dt}{\raisebox{2pt}{\tikz{\draw[black,dotted,line width=0.9pt](0,0) -- (4.4mm,0);}}}
\newcommand{\solid}{\raisebox{2pt}{\tikz{\draw[black,solid,line width=0.9pt](0,0) -- (4.4mm,0);}}}

\begin{figure}
    \centering
    \includegraphics[width=\textwidth]{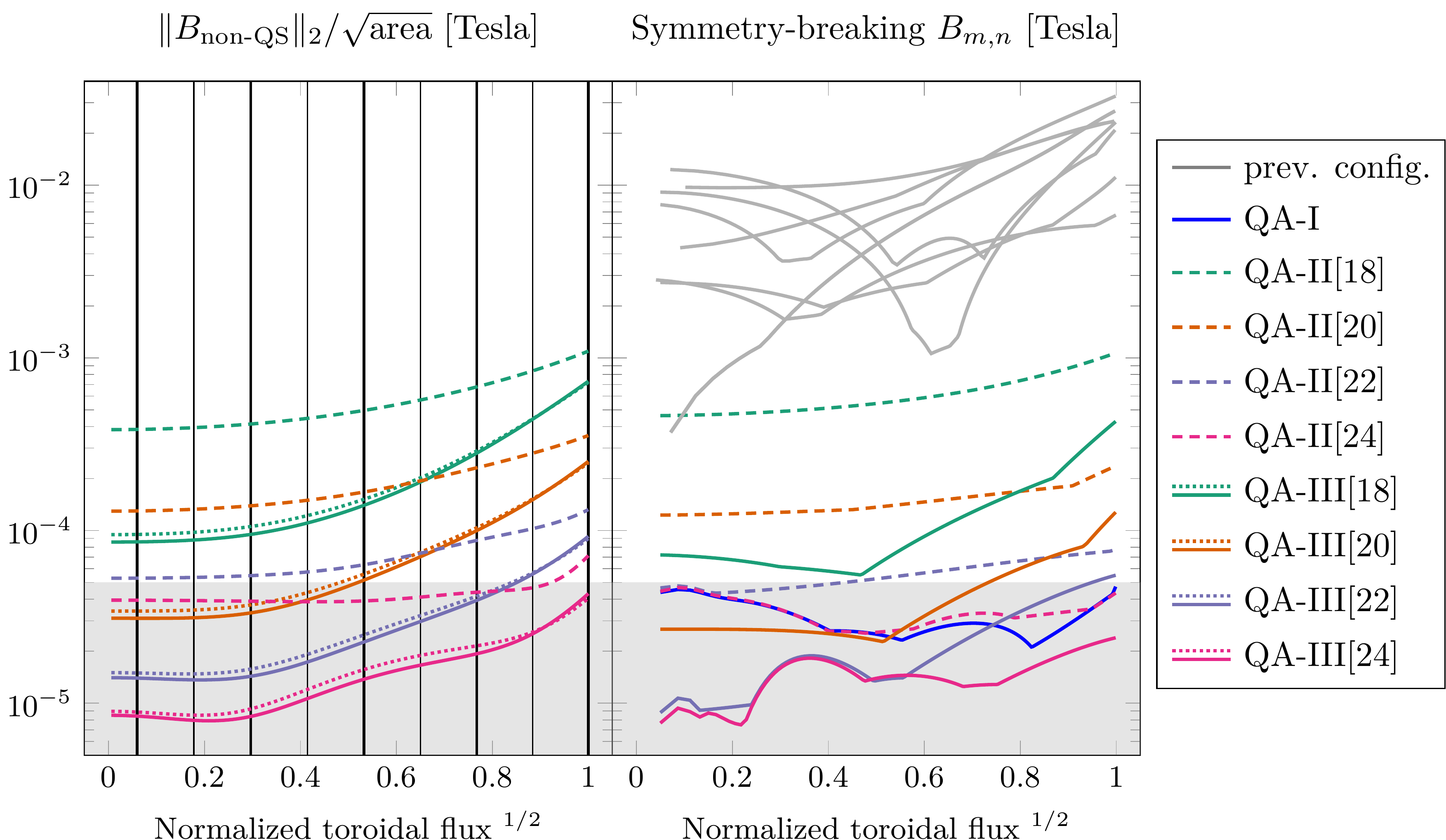}
    \caption{
    Two measures of the field's departure from quasi-symmetry: the ratio $\|B_{\mathrm{non-QS}}\|/\sqrt{\mathrm{area}}$ (left), and the maximum symmetry-breaking Fourier mode (right).  
    In both plots, the magnetic field is normalized to be $\SI{1}{\tesla}$ on the magnetic axis.
    The curves corresponding to the QA-III coils in the left figure have two line styles: dotted ({\protect\dt}), and solid ({\protect\solid}), which correspond to configurations resulting from optimization on five and nine surfaces respectively.
    The maximum symmetry-breaking Fourier mode is calculated by computing a two-dimensional FFT of the field strength, and finding the maximum $|B_{m,n}|$ for $n\neq0$, corresponding to the Fourier modes introducing variation in $\varphi$. 
    The thick and thin vertical black lines correspond respectively to the five surfaces used in the initial optimization and the four surfaces added in the nine surface optimization run.
    The shaded gray region of the plot corresponds to field strengths below \SI{50}{\micro\tesla}, corresponding to the Earth's background magnetic field.
    }
    \label{fig:ratio_and_bmn}
\end{figure}

We first optimize coils using five BoozerExact surfaces uniformly spaced in terms of the surface label $\sqrt V$, which is proportional to the minor radius. We choose $m_{\text{pol}} = n_{\text{tor}} = 10$, and the outermost surface in the initial configuration has aspect ratio approximately equal to 6, similar to the optimization in \cite{LandremanPrecise}.
We set the target toroidal flux (equation  \eqref{eq:optprob:4}) and major radius (equation \eqref{eq:optprob:5}) of the innermost surface to be the same as the one in the stage II coil set.  In addition, we use the same coil engineering requirements (equations \eqref{eq:optprob:6}-\eqref{eq:optprob:9}) as the ones used in \cite{wechsung2022precise}.
Each of the four base modular coils is described by a Fourier representation with 16 Fourier modes, which results in an optimization problem with 400 degrees of freedom corresponding to the geometry and coil currents. 

Unlike in \cite{LandremanPrecise}, we do not directly fix the aspect ratio of the outermost surface. Instead, we fix the major radius of the innermost surface and the volume of the outermost surface. We empirically observe that this effectively limits the variation of the aspect ratio of the outermost surface over the course of the optimization.
Once the coils are optimized on five surfaces, we increase the number of BoozerExact surfaces to nine by keeping the original five surfaces and introducing four new surfaces in between, and again run at most 45,000 BFGS iterations.

In the following description of the results, we will refer to the stage I precise QA magnetic field of \cite{LandremanPrecise} as QA-I, the configurations obtained from the classical stage II coil optimization in~\cite{wechsung2022precise} for different maximum coil lengths as QA-II[$L_{\max}$], and the stage III coils obtained from solving the optimization problem described in this manuscript as QA-III[$L_{\max}$].

When optimizing on nine surfaces, all coil sets considered in this manuscript reached optimality, ($\| \nabla J \|_{\infty} < 10^{-10}$), within the maximum number of iterations and the gradient was reduced by at least six orders of magnitude.
More iterations are required for longer coil lengths since the optimization problem becomes more ill-conditioned (see Figure \ref{fig:five_gradient}).
In Figure~\ref{fig:ratio_and_bmn}, we provide two similar measures of quasi-symmetry: the norm of the non-quasisymmetric component of the field magnitude and the maximum symmetry-breaking Fourier modes. 
Significant improvements in quasi-symmetry are observed after optimization on five surfaces. Additional minor improvement is observed when increasing from five to nine surfaces (Figure~\ref{fig:ratio_and_bmn}a).  This suggests that for this particular setup, five surfaces were sufficient to optimize for QA throughout the volume and not just in the neighborhood of the surfaces. 

In our implementation, we use MPI-based parallelism to launch $N_s$ tasks that concurrently compute surfaces and adjoint variables, meaning that the time taken to evaluate the objective and its gradient does not substantially increase with the number of surfaces on which quasi-symmetry is optimized. 
Since five surfaces might not be sufficient in all configurations, the possibility of parallelizing the surface and adjoint solves, as available in our implementation, means that we do not have to compromise on the number of surfaces used.
Our implementation relies heavily on the open-source stellarator optimization package SIMSOPT \citep{simsopt} where many aspects of this work are contributed.
We also use SIMSOPT's ``optimizable'' framework for efficient representation of our objective (appendix \ref{sec:optimizable}).
In the above example, a typical iteration of BFGS takes approximately two seconds on Intel Xeon Platinum 8268 processors when optimizing on five or nine surfaces.

In Figure~\ref{fig:ratio_and_bmn}b, we provide the magnitude of the maximum symmetry-breaking mode of the field magnitude and compare with eight previous configurations taken from figure 6a in \citep{LandremanPrecise}.
The shaded area of the figure indicates field magnitudes that are smaller than the Earth's background magnetic field (\SI{50}{\micro\tesla}).  
We highlight that after optimization, the symmetry-breaking errors of the QA-III[24] coils improve not only on the QA-II[24] field, but even on the original QA-I field.
This might have a number of explanations.  For example, the dimension of the optimization problem here is 400 and thus just over three times larger than the dimension of the problem used to design QA-I, which is 120.
This also might be because we use discretely exact gradients in this work, and all coils in the nine surface optimization reached optimality, to very close precision.
This is in contrast to the QA-I field computations, where only finite difference gradients were used, thus potentially limiting the accuracy of gradients and as a result how much the objective could be decreased.

\begin{figure}
    \centering
    \includegraphics[width=0.24\linewidth,trim={7cm 0 7cm 0},clip]{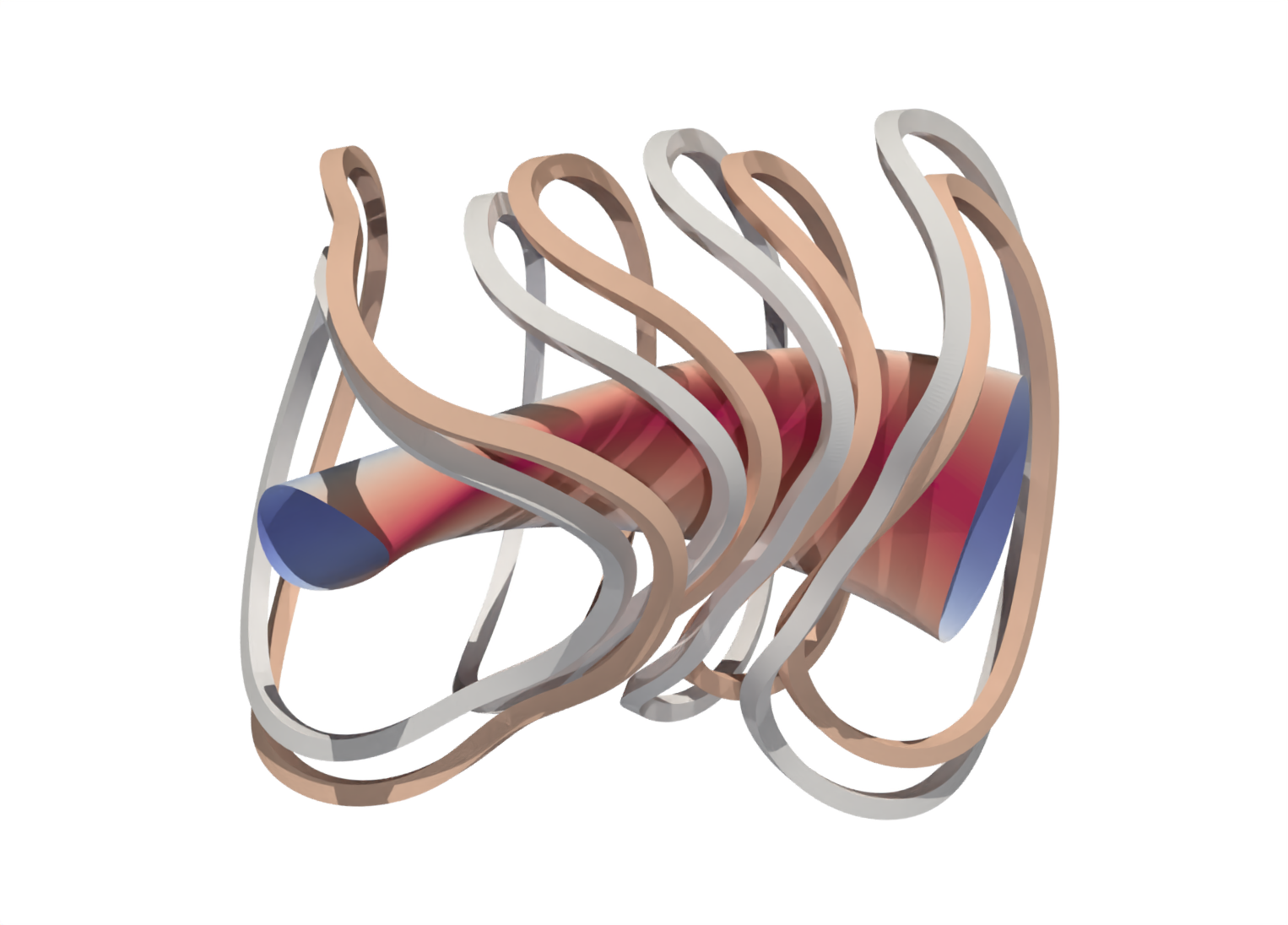}
    \includegraphics[width=0.24\linewidth,trim={7cm 0 7cm 0},clip]{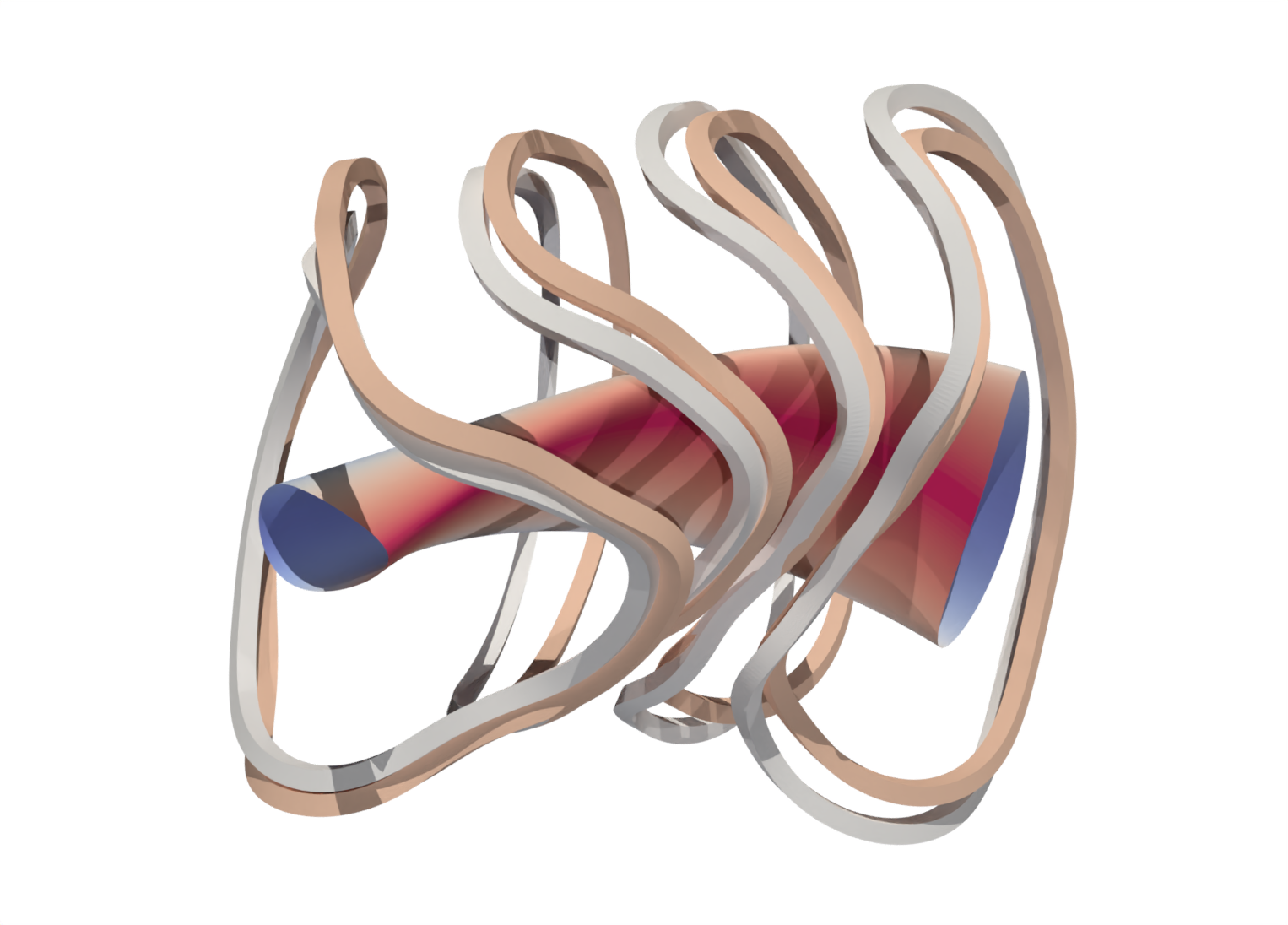}
    \includegraphics[width=0.24\linewidth,trim={7cm 0 7cm 0},clip]{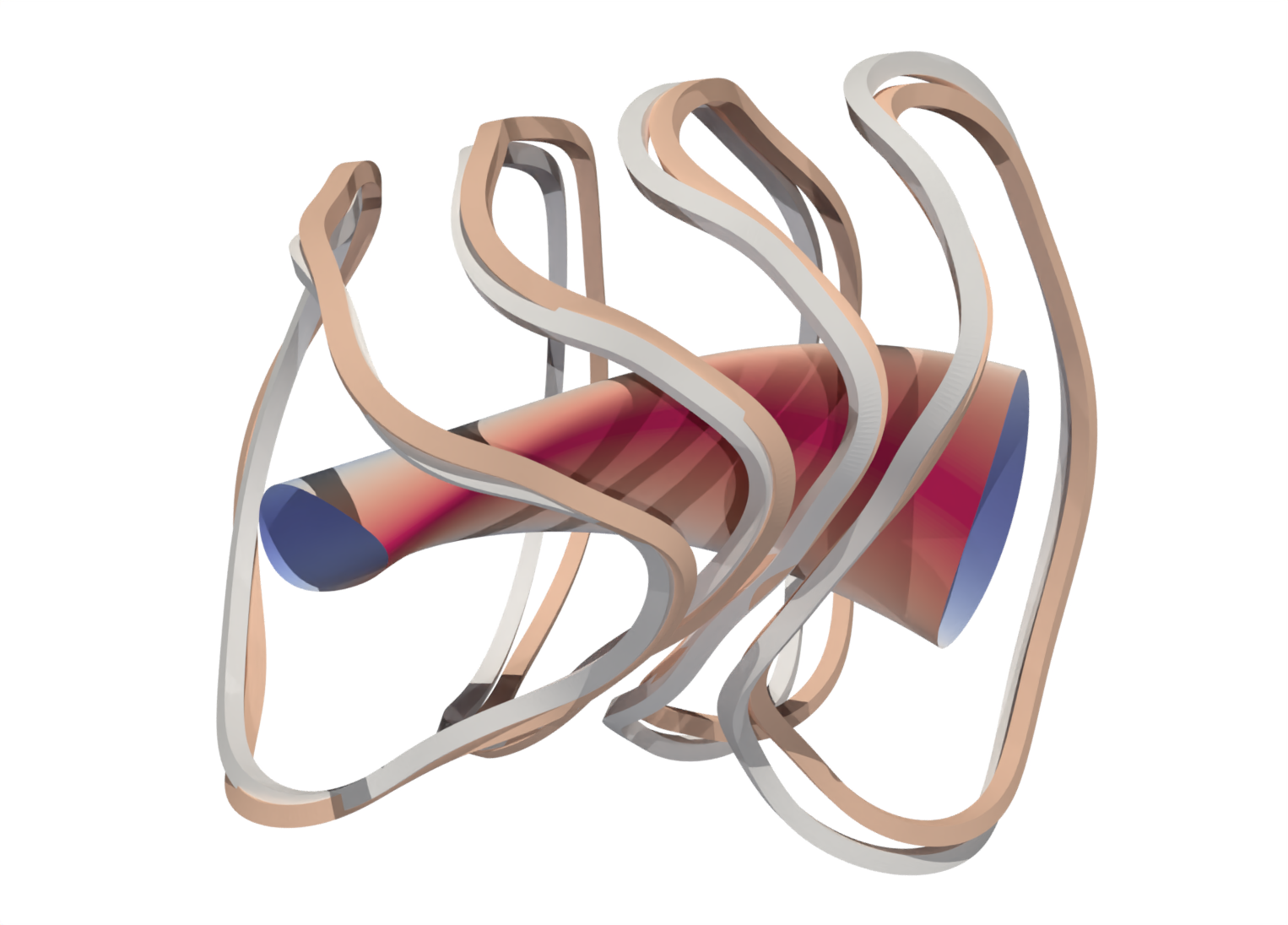}
    \includegraphics[width=0.24\linewidth,trim={7cm 0 7cm 0},clip]{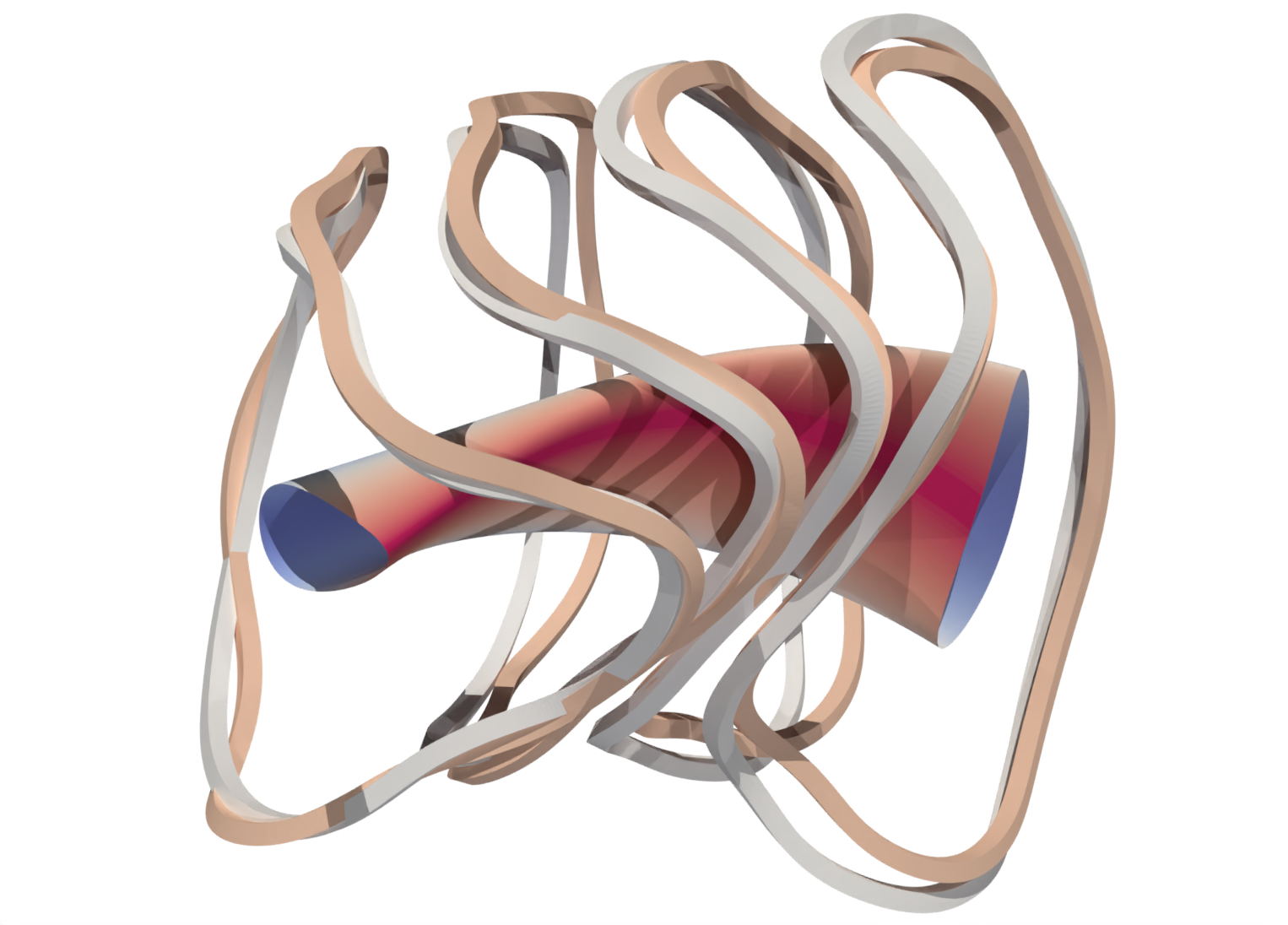}

    \caption{From left to right, inboard view of the $\SI{18}{\meter}, \SI{20}{\meter}, \SI{22}{\meter}, \SI{24}{\meter}$ coils before and after optimization. Grey and gold correspond respectively to the stage II coils and stage III coils (nine surface optimization).  The color on the surface corresponds to the magnetic field strength.
    }
    \label{fig:coils}
\end{figure}

In the physics results discussed below, we now only analyze the coils optimized for QA using nine surfaces.
Geometric properties of the final coils after optimization for QS on nine surfaces are provided in Table \ref{tab:geoprop} and the coils are shown in Figure~\ref{fig:coils}.
The full coil set for the QA-III[24] configuration is shown in Figure~\ref{fig:stellarator_epseff} (left).
As $L_{\max}$ increases, the coil-to-surface separation increases as coils can move further away from the outermost surface, thereby reducing the effects of the discrete nature of the coils and producing better quasi-symmetry in the volume.
As the coil length increases, the maximum curvatures, mean-squared curvatures, and the coil-coil separation approach and attain $\kappa_{\max}$, $\kappa_{\mathrm{msc}}$, $d_{\min}$, respectively.
We observe that although the surfaces do not change substantially in the optimization (Figure \ref{fig:xs}), the coils do (Figure \ref{fig:coils}).

\begin{table}\centering
    \begin{tabular}{r|p{3.5cm} p{3.5cm} p{1.9cm} p{1.9cm}}
        $L_{\max}$        & Maximum\newline curvatures & Mean-squared \newline curvatures & Coil-coil\newline separation & Coil-surf.\newline separation \\\midrule
        QA-II[18]      & $4.33,4.42,4.29,4.47$  & $4.56,4.85,4.90,4.98$ & 0.132 & 0.271\\
        QA-II[20]      & $4.89,4.84,4.67,5.00$  & $4.60,5.00,5.00,5.00$ & 0.113 & 0.287\\
        QA-II[22]      & $5.00,5.00,5.00,5.00$  & $4.17,5.00,5.00,5.00$ & 0.099 & 0.303\\
        QA-II[24]      & $5.00,5.00,5.00,5.00$  & $3.99,4.92,5.00,5.00$ & 0.099 & 0.307\\
        \midrule
        QA-III[18]      & $3.54,4.07,4.74,4.36$  &  $5.00,5.00,5.00,4.30$ &  0.108 &  0.239\\
        QA-III[20]      & $4.76,5.00,5.00,4.97$  &  $5.00,5.00,5.00,5.00$ &  0.099 &  0.272\\
        QA-III[22]      & $4.87,5.00,5.00,5.00$  &  $5.00,5.00,5.00,5.00$ &  0.099 &  0.289\\
        QA-III[24]      &  $5.00,5.00,5.00,5.00$  &  $5.00,5.00,5.00,5.00$ & 0.099 &  0.297\\

    \end{tabular}
    \caption{Comparison of geometric coil properties obtained from stage II and stage III optimization for different modular coil lengths. Stage III results are optimized using nine surfaces.  Coil-to-surface distance is computed with respect to the surface used for coil design in the stage II rows, and the outermost surface on which quasi-symmetry is optimized for the stage III rows.}\label{tab:geoprop}
\end{table}

Cross sections of surfaces in the magnetic field after stage III optimization on nine surfaces and Poincar\'e plots are provided in Figure \ref{eq:xs_poincare} (left), where there are no visible islands.
The rotational transform is approximately constant and equal to $\overline{\iota}$ throughout the volume on which we are optimizing, varying between $0.415$ and $0.428$ for the different coil lengths.
We also compare cross sections of surfaces in the stage II and stage III fields in Figure \ref{eq:xs_poincare} (right).  We observe that although the coils change substantially (Figure \ref{fig:coils}), the surfaces do not.  

\begin{figure}
    \centering
    \includegraphics[width=\linewidth]{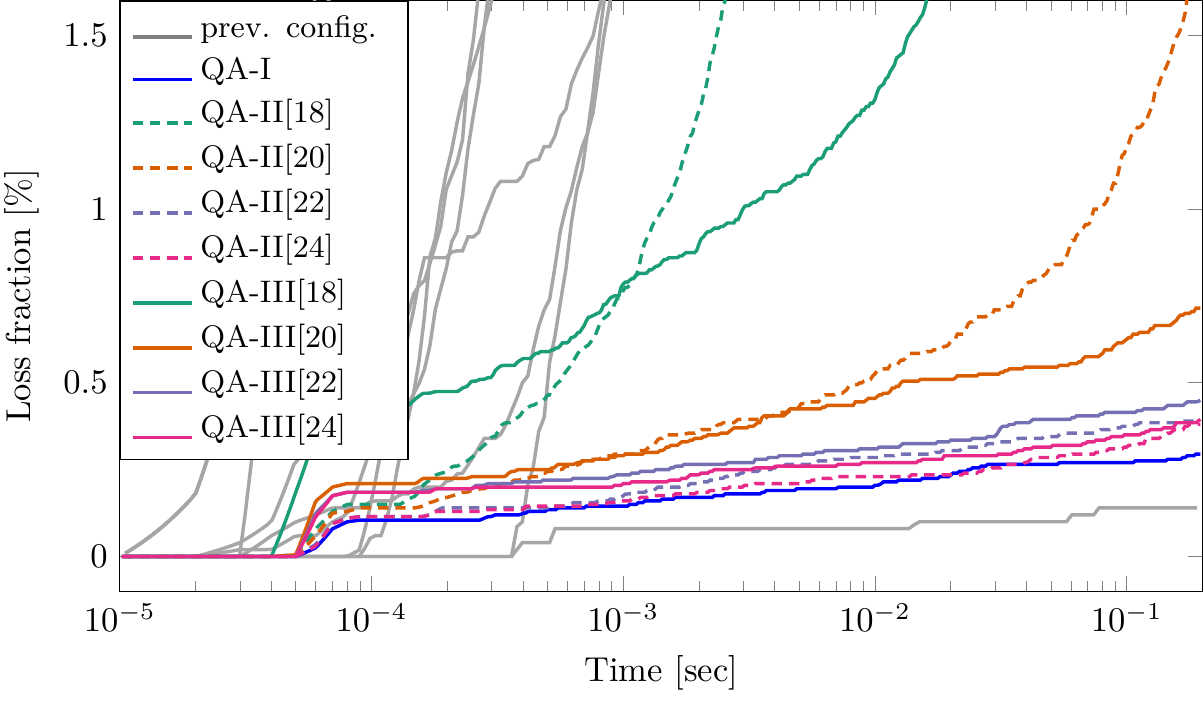}
    \caption{
    Losses of alpha particles spawned on the surface with normalized toroidal flux $0.25$. We observe significant improvements in confinement for the shorter coils.
    For $L_{\max}=\SI{18}{\meter}$, losses are reduced from 17.7\% to 6.6\%, for $L_{\max}=\SI{20}{\meter}$, losses are reduced from 1.7\% to 0.7\%.
    The stage II and III coils with lengths $L_{\max}=\SI{22}{\meter}, \SI{24}{\meter}$ present comparable particle losses, despite significantly better quasi-symmetry for the stage III coil sets.
    The previous configurations, shown in grey, are detailed in \cite{LandremanPrecise}.
}\label{fig:confinement}
\end{figure}

\begin{figure}
    \centering
    \includegraphics[height=4.5cm]{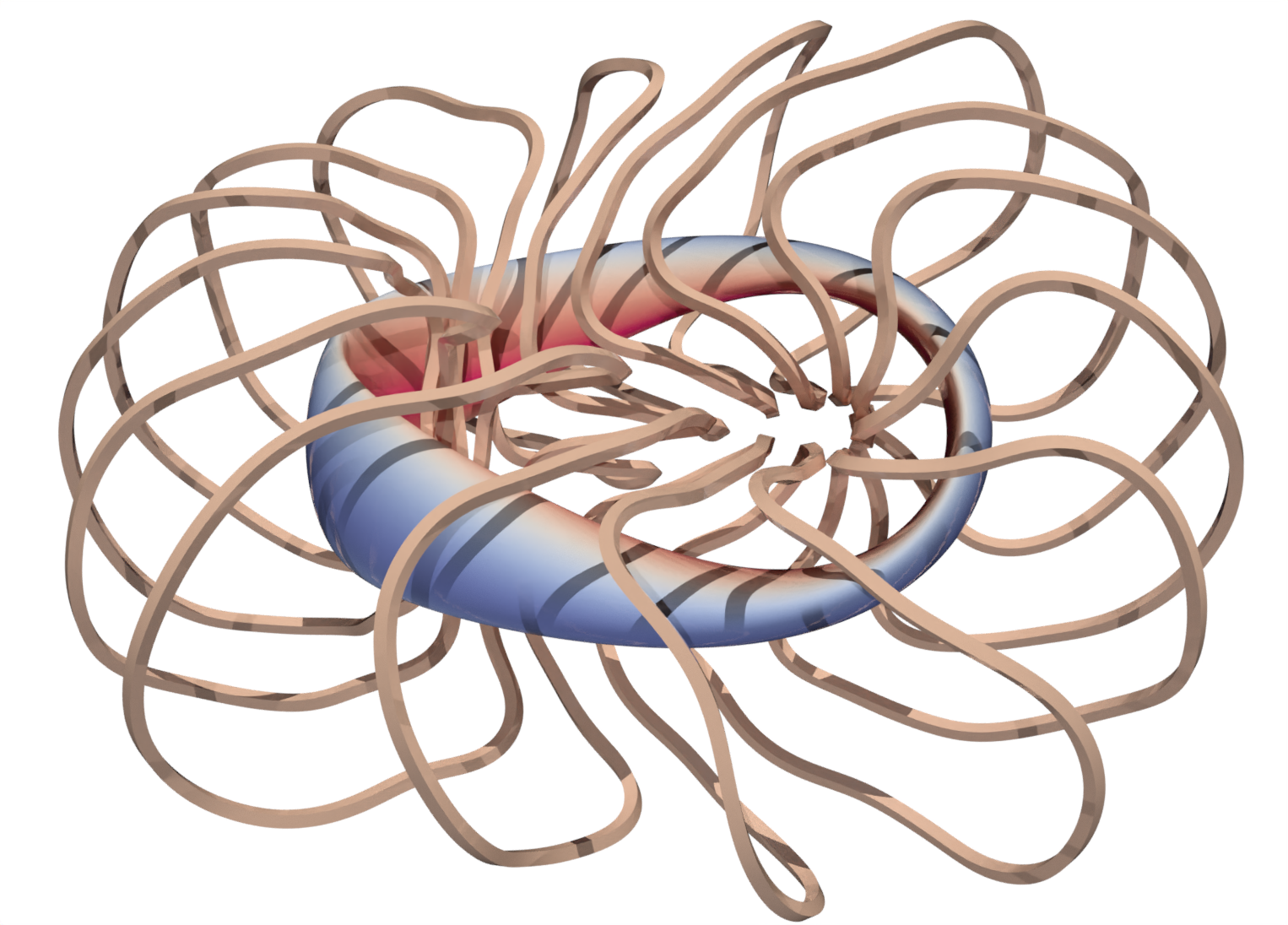} \hfill
    \includegraphics[height=4.5cm]{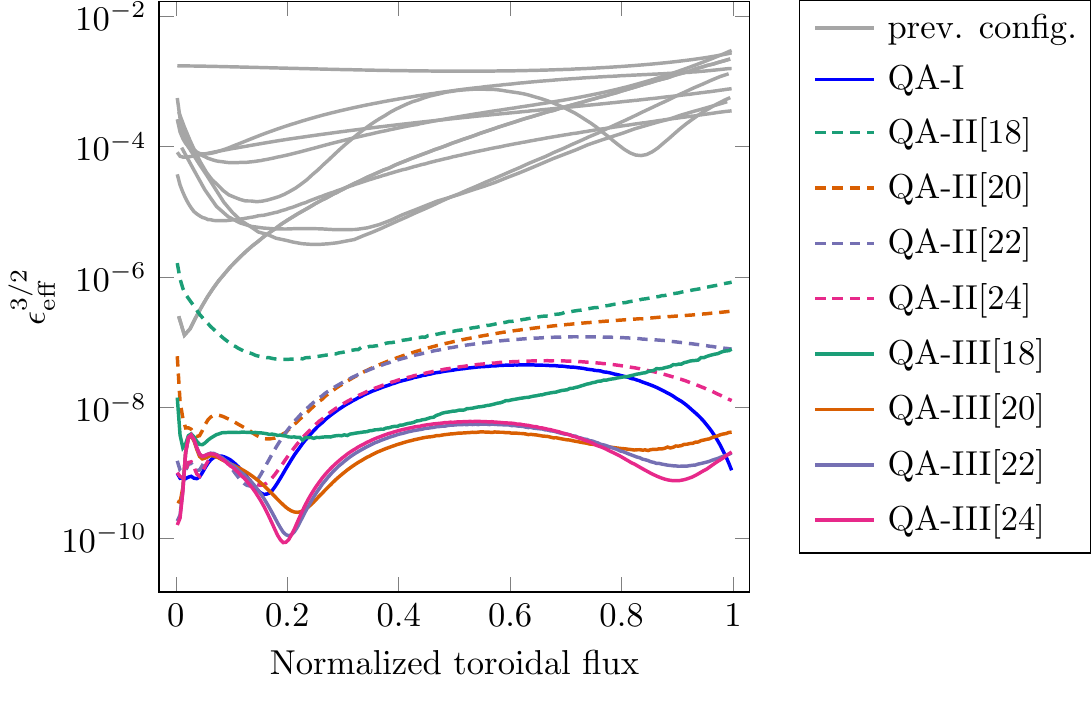}
    \caption{\textit{Left}: The QA-III[24] stellarator and the outermost surface on which quasi-symmetry is optimized.
    \textit{Right}: The thermal collisional transport magnitude $\epsilon_{\mathrm{eff}}^{3/2}$ is significantly reduced; the QA-III[20], QA-III[22], and QA-III[24] configurations even improve on the QA-I configuration.
    The previous configurations, shown in grey, are detailed in \cite{LandremanPrecise}.
}\label{fig:stellarator_epseff}
\end{figure}

\begin{figure}
    \centering
    
     \includegraphics[width=\linewidth]{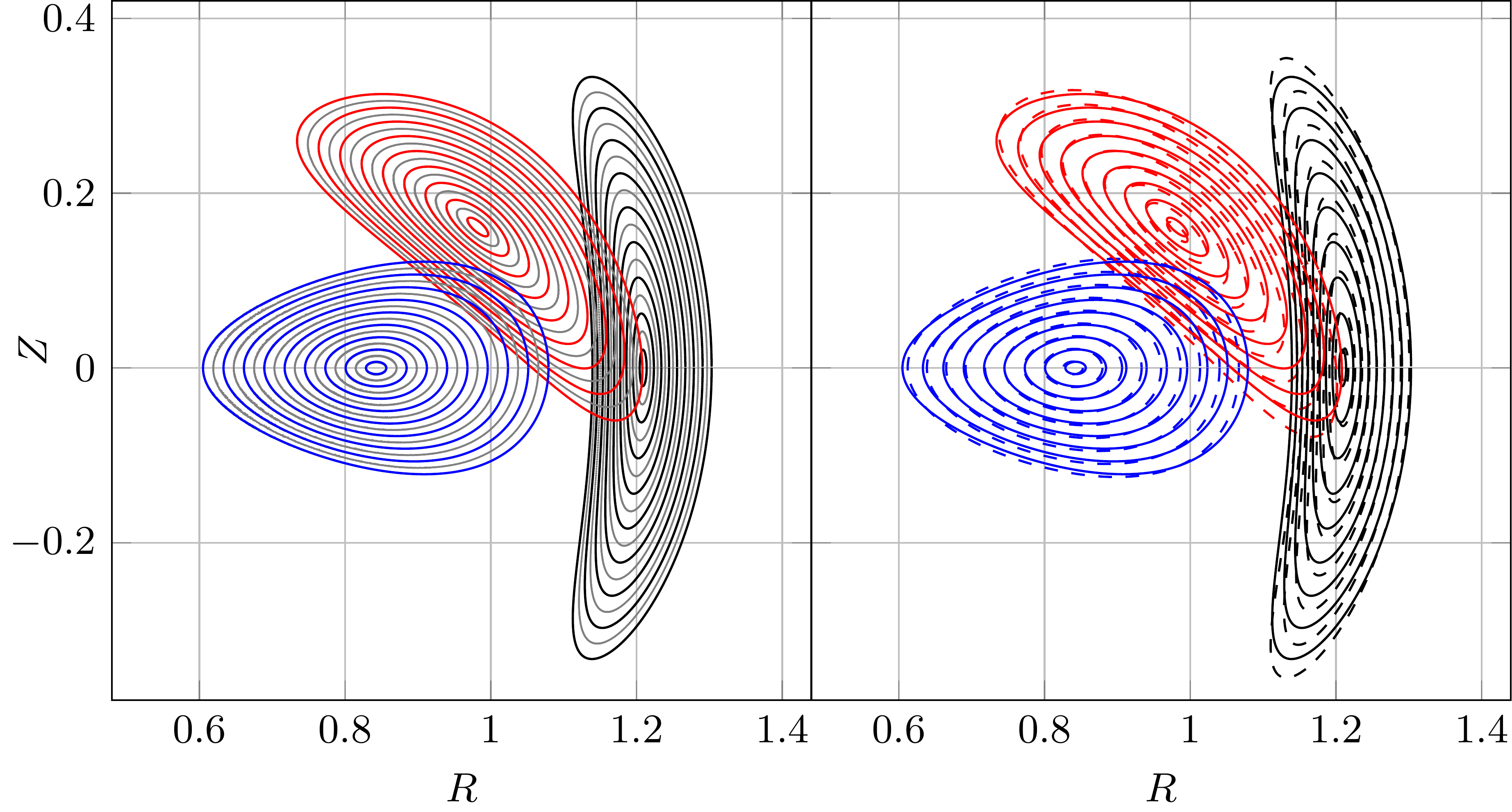}
    \caption{Shown on the left and right with solid lines ({\protect\solid}) are cross sections of the nine surfaces used in the QA-III[18] coil set optimization at $\phi = 0, \pi/4, \pi/2$ in black, red and blue. 
    Grey lines on the left correspond to Poincare plots.
    Shown on the right are cross sections of surfaces in the QA-II[18] field using dashed lines (\protect\ds). The longer coil sets result in comparable cross sections, but the difference between stage II and III cross sections become smaller.}\label{eq:xs_poincare}
    \label{fig:xs}
\end{figure}

We compute two measures to quantify the confinement properties of our optimized magnetic configurations: alpha particle losses (Figure~\ref{fig:confinement}) and the thermal collisional transport magnitude $\epsilon_{\mathrm{eff}}^{3/2}$ (Figure~\ref{fig:stellarator_epseff} right).
The alpha particle losses are calculated as follows. Each configuration is first scaled to match the minor radius and average field strength of the ARIES-CS reactor, \SI{1.70}{\meter} and \SI{5.9}{\tesla}. Alpha particles are initialized isotropically in velocity space with 3.5 MeV energy, and uniformly on the flux surface with toroidal flux $1/4$ that of the outermost computational surface, corresponding to half of the effective minor radius. The collisionless guiding-center trajectories are followed using the SIMPLE code \citep{AlbertJPP, AlbertJCP}, and particles are considered lost if they exit the outermost surface.
When considering the confinement of alpha particles, we see the most significant improvements for the two shorter coil sets.
For QA-III[18], losses are reduced from 17.7\% to 6.6\%, for QA-III[20], losses are reduced from 1.7\% to 0.7\%.
For the longer QA-III[22] and QA-III[24] coils, the (small) particle losses are comparable to those for QA-II[22] and QA-II[24] respectively, despite significantly better quasi-symmetry throughout the volume.
We note that this imperfect correlation between quasi-symmetry and energetic particle confinement is consistent with the findings in \cite{BaderIAEA}, and in \cite{LandremanPrecise} where the QA+Well configuration had lower particle losses than the QA configuration despite worse quasi-symmetry.
For $\epsilon_{\mathrm{eff}}^{3/2}$ \citep{NEO} we observe significant improvement compared to the coils of \cite{wechsung2022precise}, and our procedure is even able to improve the results from the target field of \cite{LandremanPrecise}.

\section{Conclusions}
The main contributions of this work are twofold: the introduction of a novel approach for computing surfaces directly parametrized in Boozer coordinates (section \ref{sec:surfaces}), and the optimization of a vacuum magnetic field generated by coils for quasi-symmetry on a number of magnetic surfaces (section \ref{sec:qs}).

The method we call ``BoozerExact surfaces'' can be used to accurately compute magnetic surfaces when islands or chaos are not present.
In contrast, we have shown that the method we call ``BoozerLS surfaces'' is a more robust approach to compute surfaces, even in regions where nested flux surfaces do not exist. In addition, the BoozerLS residual can effectively be used to identify regions where the assumption of nested flux surfaces is no longer valid.

In the second part of the article, we have shown that our new framework to compute magnetic surfaces can be used to optimize the geometry and currents of filamentary coils, with the goal of improving the quasi-symmetry of the magnetic field generated by these coils. At the moment, this approach requires an initial coil set that already produces a magnetic field with surfaces that can be parametrized in Boozer coordinates. In other words, it requires a ``warm-start''.  In addition, the method only works in a robust manner if islands or chaotic regions do not appear over the course of the coil optimization.  In section \ref{sec:opt-results}, we demonstrate that this technique is highly effective when these requirements are satisfied.
In fact, for the QA-III[24] configuration, we found a coil design with quasi-symmetry that improved on the quasi-symmetry of the original QA-I configuration.
This could be attributed to the higher dimensionality of our optimization problem, or our use of analytical gradients as opposed to those obtained using finite differences.
For the shorter coil configuration QA-III[18], we are able to significantly improve performance compared to coils obtained from the standard two-stage approach: the thermal collisional transport magnitude $\epsilon^{3/2}_{\text{eff}}$ is reduced by more than an order of magnitude, and alpha particle losses are reduced from $17.7\%$ to $6.6\%$. Future work will include using BoozerLS surfaces and their residual to optimize coils with less suitable initial coil configurations.
We also will consider a stochastic version of this objective to find coils that are robust to manufacturing errors.

\section*{Availability of code and optimized configurations}
The coil and surface parametrizations, Biot-Savart kernel, as well as the solver used to compute BoozerExact and BoozerLS surfaces 
were implemented in the SIMSOPT package \citep{simsopt}, 
available at:
\begin{center}
\url{https://github.com/hiddenSymmetries/simsopt}.    
\end{center}
Driver scripts for work are available at:
\begin{center}
\url{https://github.com/andrewgiuliani/PySurfaceOpt}.    
\end{center}
The stage III coils discussed in this work are available at the above repository as well.

\section*{Acknowledgements}
The authors would like to thank the SIMSOPT development team. This work was supported by a grant from the Simons Foundation (560651). AG is partially supported by an NSERC (Natural Sciences and Engineering Research Council of Canada) postdoctoral fellowship.  In addition, AC and FW are  supported by the United States National Science Foundation under grant No. PHY-1820852, and AC is supported by the United States Department of Energy, Office of Fusion Energy Sciences, under grant No. DE-FG02-86ER53223.

\FloatBarrier

\bibliographystyle{jpp}
\bibliography{references}

\begin{appendices}
\section{Properties of the surface discretization}\label{SurfaceRot}
We give a brief derivation of the surface discretization described in Section~\ref{sec:surface-param}.
First, note that any periodic smooth function $f$ in one dimension can be approximated arbitrarily well by functions of the form
\begin{equation*}
    f_{\ntor}(\varphi) = \sum_{i=1}^{2\ntor+1} a_i v_i(\varphi), v_i \in V^{\ntor}
\end{equation*}
where $V^{\ntor} = \{1, \cos(2\pi \varphi), \ldots, \cos(\ntor 2\pi \varphi), \sin(2\pi \varphi), \ldots, \sin(\ntor 2\pi \varphi)\}$.
Hence, by considering the tensor product of this Fourier basis, we obtain that any periodic smooth function $g$ of two variables can be approximated arbitrarily well by functions of the form
\begin{equation*}
    g_{\ntor, \mpol} (\varphi, \theta) = \sum_{i=1}^{2\ntor+1}\sum_{j=1}^{2\mpol+1} a_{i,j} v_i(\varphi)w_j(\theta), v_i \in V^{\ntor},w_j\in V^{\mpol}.
\end{equation*}
Hence, any surface $\bm \Sigma(\varphi, \theta)$ that can be smoothly parametrized in terms of periodic angles $\varphi$ and $\theta$ can be approximated by
\begin{equation*}
    \bm \Sigma(\varphi, \theta) =
    \begin{bmatrix}
        \sum_{i,j} x_{i,j}v_i(\varphi)w_j(\theta)\\
        \sum_{i,j} y_{i,j}v_i(\varphi)w_j(\theta)\\
        \sum_{i,j} z_{i,j}v_i(\varphi)w_j(\theta)\\
    \end{bmatrix},
\end{equation*}
where the bounds on $i$ and $j$ in the double summation have been neglected for brevity.
To describe stellarator relevant surfaces efficiently, we want to be able to exploit rotational symmetry.
Denoting the rotation matrix by
\begin{equation*}
    M(\varphi) = \begin{bmatrix}
         \cos(2\pi\varphi) & - \sin(2\pi\varphi) & 0\\
         \sin(2\pi\varphi) & ~~~\sin(2\pi\varphi) & 0\\
        0 & 0 & 1
    \end{bmatrix},
\end{equation*}
a surface $\bm \Sigma$ satisfies $\nfp$-fold rotational symmetry if
\begin{equation*}
    \bm \Sigma(\varphi+1/\nfp, \theta) = M(1/\nfp) \bm \Sigma(\varphi, \theta).
\end{equation*}
Hence, denoting
\begin{equation*}
    \tilde{\bm \Sigma}(\varphi,\theta) = M(-\varphi) \bm \Sigma(\varphi, \theta),
\end{equation*}
we note that $\tilde{\bm \Sigma}$ is $1/\nfp$ periodic in $\varphi$, since
\begin{equation*}
    \tilde{\bm \Sigma}(\varphi+1/\nfp,\theta) = M(-\varphi-1/\nfp) \bm \Sigma(\varphi+1/\nfp, \theta) = M(-\varphi) \bm \Sigma(\varphi, \theta) = \tilde{\bm \Sigma}(\varphi, \theta).
\end{equation*}
We conclude that a surface with $\nfp$-fold rotational symmetry can be approximated by functions of the form
\begin{equation*}
    \bm \Sigma(\varphi, \theta) = M(\varphi)
    \begin{bmatrix}
        \sum_{i,j} x_{i,j}v_i(\nfp\varphi)w_j(\theta)\\
        \sum_{i,j} y_{i,j}v_i(\nfp\varphi)w_j(\theta)\\
        \sum_{i,j} z_{i,j}v_i(\nfp\varphi)w_j(\theta)\\
    \end{bmatrix}
\end{equation*}

\section{Derivation of the major radius formula}\label{Appen:Major}
The formula for the major radius is computed from the mean cross sectional area $\bar A$ and volume enclosed by the surface
$$
R_{\mathrm{major}}(\mathbf{s}) =  V(\mathbf{s}) / (2 \pi \bar A(\mathbf s)),
$$
where 
$$
\bar A(\mathbf{s}) = \frac{1}{2 \pi}\int^{\pi}_{-\pi} A_{\phi} ~d\phi,
$$
where $A_{\phi}$ is the surface's cross sectional area computed at the cylindrical angle $\phi$.  Note that there are multiple ways to evaluate this term. One possibility is to determine the cross sectional area at a fixed number of cylindrical angles $\phi$, and then average.  This is difficult with our surface representation, as it requires solving several nonlinear equations. We follow an alternative approach that is less complex implement and more elegant.  The cross sectional area can be written as
$$
A_{\phi} = \int^1_{0} \tilde R \frac{\partial \tilde Z}{\partial \theta} ~ d\theta,
$$
where $\tilde R(\theta), \tilde Z(\theta)$ are points on the boundary of the cross section at the angle $\phi$.  Next, writing the above integrand in terms of the surface's Boozer angles, we have
$$
A_{\phi} = \int^1_0 R(\varphi(\phi, \theta), \theta) \frac{\partial }{\partial \theta}[Z(\varphi(\phi, \theta), \theta)] ~d\theta,
$$
where $ \tilde R = R(\varphi(\phi, \theta), \theta)$ and  $ \tilde Z = Z(\varphi(\phi, \theta), \theta)$.
The expression for the average cross sectional area then becomes
$$
\bar A = \frac{1}{2 \pi}\int^{\pi}_{-\pi}  \int^1_0 R(\varphi(\phi, \theta), \theta) \frac{\partial }{\partial \theta}[Z(\varphi(\phi, \theta), \theta)] ~d\theta ~d\phi,
$$
Instead of integrating over cylindrical $\phi$, we complete the change of variables
\begin{equation*}
    (\phi, \theta) = (\text{atan2}(y(\varphi, \theta), x(\varphi, \theta)), \theta).
\end{equation*}
After the change of variables, the average cross sectional area is
$$
\bar A = \frac{1}{2 \pi}\int^{1}_0  \int^1_0 R(\varphi, \theta) \left[ \frac{\partial Z}{\partial \varphi}\frac{\partial \varphi}{\partial \theta} + \frac{\partial Z}{\partial \theta}\right]~ \text{det}~J ~d\varphi~d\theta ,
$$
where $\text{det}~J$ is the determinant of the mapping's Jacobian.

\section{Optimizable graph for efficient representation of the objective}\label{sec:optimizable}
\begin{figure}
    \centering
    \includegraphics[width=0.6\textwidth]{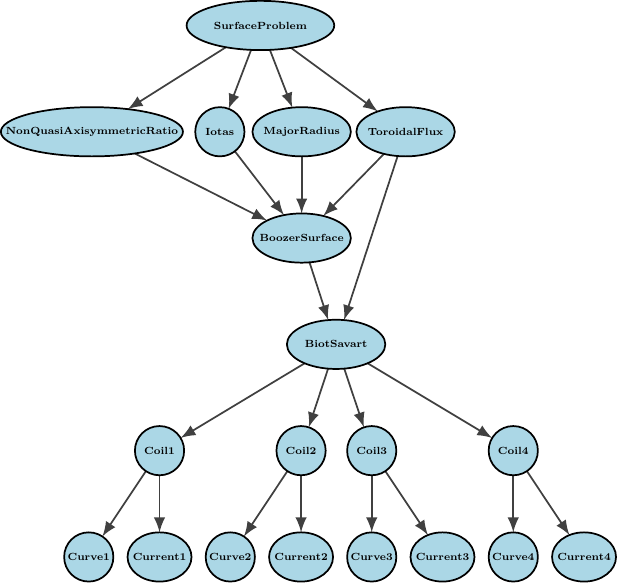}
    \caption{A simplified dependency graph of the various quantities used to define the optimization problem.  In this model problem, there is one surface (``BoozerSurface'' node) on which we are trying to optimize for quasi-symmetry (``NonQuasiAxisymmetricRatio'' node), target a specific rotational transform (``Iotas'' node), major radius (``MajorRadius'' node) and toroidal flux (``ToroidalFlux'' node).  
These penalty terms depend on on the magnetic field computed via the Biot Savart law (``BiotSavart'' node) and on the geometry of the Boozer surface (``BoozerSurface'' node).
The Biot Savart node then depends directly on the four stellarator coil geometries and associated currents.}
    \label{fig:opt_graph}
\end{figure}
The different penalty terms of the optimization problem in section \ref{sec:opt} are implemented using the ``optimizable'' framework in SIMSOPT \citep{simsopt}.
The dependencies between the various quantities in optimization problems result in a directed acyclical graph, which can be used for efficient caching, i.e., terms are only recomputed when quantities that they depend on are changed.
The dependency graph for a simplified version of the optimization problem solved in section \ref{sec:opt} is provided in Figure \ref{fig:opt_graph}.
A quantity associated to any of the nodes in the graph is only recomputed if any of its ancestors have been modified.
The full optimization problem solved in section \ref{sec:opt} results in a much more complex dependency graph that we do not show here for brevity.

\end{appendices}
\end{document}